\documentclass[manuscript]{aastex}

\pdfoutput=1
\usepackage{amsmath,amssymb}
\usepackage{bm}
\usepackage[usenames]{color}

\newcommand{\major}{black}

\newcommand{\add}{black}
\newcommand{\minor}{black}
\newcommand{\oldmajor}{black}
\newcommand{\oldminor}{black}
\newcommand{\sgn}{\operatorname{sgn}}
\newcommand{\conv}{adv}

\slugcomment{}

\shorttitle{Convection Enhances Magneto-rotational Turbulence}
\shortauthors{Hirose et al.}

\begin{document}


\title{Convection Causes Enhanced Magnetic Turbulence \\
in Accretion Disks in Outburst}


\author{Shigenobu Hirose}
\affil{Institute for Research on Earth Evolution, Japan Agency for Marine-Earth Science and Technology, Yokohama, Kanagawa 236-0001, Japan}
\email{shirose@jamstec.go.jp}

\author{Omer Blaes}
\affil{Department of Physics, University of California, Santa Barbara, CA 93106}

\author{Julian H. Krolik}
\affil{Department of Physics and Astronomy, Johns Hopkins University, Baltimore, MD 21218}

\author{Matthew S. B. Coleman}
\affil{Department of Physics, University of California, Santa Barbara, CA 93106}

\and

\author{Takayoshi Sano}
\affil{Institute of Laser Engineering, Osaka University, Suita, Osaka 565-0871, Japan}




\begin{abstract}
We present the results of local, vertically stratified, radiation magnetohydrodynamic (MHD) shearing
box simulations of magneto-rotational (MRI) turbulence appropriate for the hydrogen ionizing regime
of dwarf nova and soft X-ray transient outbursts.  We incorporate the
frequency-integrated opacities and equation of state for this
regime, but neglect non-ideal MHD effects and surface irradiation, and do
not impose net vertical magnetic flux.  We find two stable thermal equilibrium
tracks in the effective temperature versus surface mass density plane,
in qualitative agreement with the S-curve picture of the standard disk
instability model.  We find that the large
opacity at temperatures near $10^4$K,
a corollary of the hydrogen ionization transition, 
triggers strong, intermittent thermal convection on the
upper stable branch.  This convection strengthens the magnetic turbulent dynamo and greatly enhances
the time-averaged value of the stress to thermal pressure ratio $\alpha$, {\color{\oldmajor}possibly}
by generating vertical magnetic field that may seed the axisymmetric MRI, and by increasing cooling
so that the pressure does not rise in proportion to the turbulent dissipation.
These enhanced stress to pressure ratios
may alleviate the order of magnitude discrepancy between the $\alpha$-values
observationally inferred in the outburst state 
and those that have been measured
from previous local numerical simulations of magnetorotational
turbulence that lack net vertical magnetic flux.
\end{abstract}





\section{Introduction}

The accretion of material through a rotationally supported disk orbiting a
central gravitating body is a process of fundamental astrophysical importance.
In order for material to move inward through the disk to liberate 
its gravitational energy, its angular
momentum must be extracted, so that the material loses its rotational support
against gravity. The fluid stresses responsible for these torques are therefore
central to the accretion disk phenomenon.
Theoretical models of accretion disks that have been used to fit real data
generally parameterize the stresses by a dimensionless parameter $\alpha$,
the stress measured in terms of local thermal pressure \citep{Shakura_73}. 
The most reliable estimates of $\alpha$ come from episodic outbursts in dwarf
novae.  The outburst cycle in these systems is very successfully modeled by
disk instability models (DIMs)
\citep[][for a recent review, see \citet{Lasota_01}]{Osaki_74,Hoshi_79,Meyer_81,Cannizzo_82,Faulkner_83,Mineshige_83}
as a limit cycle between {\color{\major}two stable} 
thermal
equilibrium states: in outburst (high mass accretion rate) a hot state in which
hydrogen is fully ionized, and in quiescence (low mass accretion rate) a cool
state in which hydrogen is largely neutral.
The measured
outburst time scales give well-determined estimates of $\alpha\sim 0.1$ in the hot, ionized state \citep[e.g.][]{Smak_99}. {\color{\major}On the other hand, measured time intervals between outbursts indicate that $\alpha$ in the cool state is an order of magnitude smaller \citep[e.g.][]{Cannizzo_88}.}

A plausible physical mechanism for the stresses in
ionized disks is correlated magnetohydrodynamic (MHD) turbulence stirred
by nonlinear development of the magneto-rotational instability (MRI)
\citep{Balbus_91}. The MRI grows because magnetic fields in an electrically
conducting plasma cause angular momentum exchange between fluid elements that
taps the free energy of orbital shear \citep{Balbus_98}. However, numerical
simulations of this turbulence within local patches of accretion disks so far
show a universal value of $\alpha\sim 0.01$ unless net vertical magnetic flux is
imposed from the outside \citep{Hawley_95,Hawley_96,Sano_04,Pessah_07}. This
value is an order of magnitude smaller than the value suggested by the
observations of ionized outbursting disks in dwarf novae \citep{King_07}. 

It is possible that in real disks, local net flux is created by global linkages
\citep{Sorathia_10}.  However,
the centrality of the hydrogen ionization transition to DIMs of dwarf novae may
be a clue to the apparent discrepancy in $\alpha$.  A sharp change in ionization
can alter the opacity and equation of state (EOS) of a fluid, with dynamical
consequences if convection arises.  Most previous numerical studies that showed
$\alpha\sim 0.01$ assumed isothermal disks.  
In a recent attempt to understand dwarf nova disks in the framework
of MRI turbulence, \citet{Latter_12} first demonstrated the
bistability of the disk with an analytic approximate local cooling model, but without
vertical stratification and therefore without the possibility of
convection; the resultant $\alpha$ was $\sim 0.01$ in the absence of net
magnetic flux. 
To explore the
generic consequences of convection in stratified MRI turbulence, \citet{Bodo_12}
solved an energy equation with finite thermal diffusivity and a perfect gas EOS;
they found that convection enhanced the stress, 
but a notable change in $\alpha$ was not mentioned.

Here we present radiation MHD simulations that fully take into account vertical
stratification and realistic thermodynamics to determine the state of MRI
turbulence in dwarf nova disks. We include opacities and an EOS
that reflect the ionization fraction. The local thermal state is determined by a
balance between local dissipation of turbulence and cooling calculated from a
solution of the radiative transfer problem and a direct simulation of thermal
convection. We consider the case of zero net vertical magnetic field, which, as
noted above, results in the lowest possible $\alpha$ values. We assume ideal MHD in
order to focus on the effects of opacities and the EOS on the
thermal equilibrium and turbulent stresses. Non-ideal effects will likely be
very important for the cool state in which hydrogen is mostly neutral {\color{\oldmajor}\citep{Gammie_98,Sano_02,Sano_03,Kunz_13}}.

Our simulations are successful in reproducing the two distinct branches of
thermal equilibria inferred by the DIM:  a hot ionized branch and a cool
neutral branch.  We measure $\alpha$ in all our simulations and find that 
its value is significantly enhanced {\color{\oldminor}at the low surface brightness end of}
the upper branch, due to the fact
that the high opacities produce intermittent thermal convection, which enhances
the time-averaged magnetic stresses in the MRI turbulence relative to the
time-averaged thermal pressure.

We present these results in this paper, which is organized as follows. In Section \ref{sec:methods}, we describe the numerical method and the initial condition for our radiation MHD simulations. Quantitative results about thermal equilibrium and MRI turbulence in the simulations are presented in Section \ref{sec:results}. 
{\color{\oldminor}We discuss our results in Section \ref{sec:discussion},} 
and we summarize our conclusions in Section \ref{sec:conclusion}.

\section{Methods}\label{sec:methods}

We simulate a series of local patches of an accretion disk,
treating them in the vertically-stratified shearing box approximation
\citep{Hawley_95}.
These patches differ in surface mass
density $\Sigma$, but all have the same angular velocity $\Omega =
6.4\times10^{-3}$ s$^{-1}$, which corresponds
to a distance of {\color{\oldminor}$1.23\times10^{10}$} cm from a white dwarf of $0.6 M_\odot$. 
This distance is {\color{\oldminor}$\sim 14 \times$} the radius of the white dwarf
($= 0.0126 R_\odot$).\footnote{The white dwarf radius was computed assuming the mean molecular weight
per electron is two{\color{\oldminor}, appropriate for helium, carbon/oxygen,
or neon/magnesium compositions} \citep{Nauenberg_71}.}
The simulations start from a laminar flow state with a weak magnetic field and
an appropriate thermal energy content. We then evolve the MHD fluid in the box
until it either reaches approximate steady state conditions, i.e., it achieves
both hydrostatic and thermal equilibrium in a statistical sense, or it
experiences runaway heating or cooling and is unable to reach a thermal
equilibrium.  This section outlines the details and methods of our simulations.

\subsection{Basic Equations}
The basic equations for our radiation MHD simulations are
\begin{gather}
  \frac{\partial\rho}{\partial t} + \nabla\cdot(\rho\bm{v}) = 0, \\
  \frac{\partial(\rho\bm{v})}{\partial t} + \nabla\cdot(\rho\bm{v}\bm{v}) =
  -\nabla p + \frac{1}{4\pi}(\nabla\times\bm{B})\times\bm{B} + \frac{{\kappa}_\text{R}\rho}{c}\bm{F}, \label{eq:motion}\\
  \frac{\partial e}{\partial t} + \nabla\cdot(e\bm{v}) =
  -(\nabla\cdot\bm{v})p - \left(4\pi B(T) -
  cE\right){\kappa}_\text{P}\rho, \label{eq:energy_gas}\\
  \frac{\partial E}{\partial t} + \nabla\cdot(E\bm{v}) = - \nabla\bm{v}:\mathsf{P} +  \left(4\pi B(T) - cE\right){\kappa}_\text{P}\rho - \nabla\cdot\bm{F}, \label{eq:energy_rad}\\
  \frac{\partial\bm{B}}{\partial t} - \nabla\times\left(\bm{v}\times\bm{B}\right) = 0, \label{eq:induction}
\end{gather}
where $\rho$ is the gas density, $e$ the gas internal energy, {\color{\major}$p$ the gas pressure, $T$ the gas temperature}, $E$ the radiation
energy density, $\mathsf{P}$ the radiation pressure tensor, $\bm{F}$ the
radiation energy flux, $\bm{v}$ the velocity field vector, $\bm{B}$ the magnetic
field vector (in CGS emu units), $B(T) = \sigma_\text{B}T^4/\pi$ the Planck
function ($\sigma_\text{B}$, the Stefan-Boltzmann constant), and $c$ the speed
of light. We use a flux-limited diffusion approximation of the radiative
transfer, where {\color{\oldminor}the} energy flux $\bm{F}$ and pressure tensor
$\mathsf{P}$ are related to {\color{\oldminor}the} energy density $E$ as $\bm{F} =
-(c\lambda(R)/\kappa_\text{R}\rho)\nabla E$ and $\mathsf{P} =
\mathsf{f}(R)E$. Here $\lambda(R) \equiv (2+R)/(6+3R+R^2)$ is a flux limiter
with $R \equiv |\nabla E|/(\kappa_\text{R}\rho E)$, and $\mathsf{f}(R) \equiv
(1/2)(1-f(R))\mathsf{I} + (1/2)(3-f(R))\bm{n}\bm{n}$ is the Eddington tensor
with $f(R) \equiv \lambda(R) + \lambda(R)^2R^2$ and $\bm{n}\equiv\nabla E/|\nabla E|$ \citep{Turner_01}.

The EOSs, $p = p(\rho,e/\rho)$ and $T = T(\rho,e/\rho)$, are
computed from the Saha equations assuming ionization equilibrium with solar
abundances (Fig. \ref{fig:eos}). 
{\color{\oldmajor}Our method is equivalent to that described in Appendix A in \citet{Tomida_13} with the following exceptions:  (1) we consider metals and H$_2^+$ in addition to the species employed in their method and (2) we allow the orthohydrogen to parahydrogen ratio to be determined by thermal equilibrium rather than assuming the high temperature fixed ratio of 3:1.  (Neither of these changes make
significant differences.)}

The Planck-mean opacity $\kappa_\text{P}$ and
Rosseland-mean opacity $\kappa_\text{R}$ are given as a function of density $\rho$ and gas
temperature $T$ (Fig. \ref{fig:opacity}). We have combined three published opacity
tables, \citet{Semenov_03}, \citet{Ferguson_05} and the Opacity Project (OP)
\citep{Seaton_05}, to
cover the relevant temperature range for our purpose ($10^3 < T < 10^6$~K).
Different opacity tables are connected (with linear interpolation) where they seem
mostly consistent; specifically, the Semenov et~al. and Ferguson et~al. opacities are connected
at the dust sublimation temperatures while the Ferguson et~al. and OP opacities are connected
at $T = 10^{3.7}$~K.
{\color{\add} The combined opacity tables have upper and lower bounds, beyond which the opacities are
  extended using a zero-gradient extrapolation, which is not a bad approximation for $T > 10^3$~K \citep[c.f. Figure 1 in][]{Malygin_13}.}

\subsection{Shearing Box}
In the shearing box approximation, a local patch of an accretion disk is modeled
as a co-rotating Cartesian frame $(x,y,z)$ with linearized Keplerian shear flow
$-(3/2)\Omega x$, where the $x$, $y$, and $z$ directions correspond to the
radial, azimuthal, and vertical directions, respectively \citep{Hawley_95}. In
this approximation, the inertial force terms $-2\Omega\hat{\bm{z}}\times\bm{v} +3\Omega^2x\hat{\bm{x}} - \Omega^2z\hat{\bm{z}}$ (Coriolis force + tidal force + vertical
component of gravitational force, respectively) are added to the equation of
motion (\ref{eq:motion}), where $\hat{\bm{x}}$ and $\hat{\bm{z}}$ are the unit vectors in the $x$ and $z$ direction, respectively.  Shearing-periodic, periodic, and outflow boundary conditions are
used for the $x$, $y$, and $z$ boundaries of the box, respectively \citep{Hirose_06}.

\subsection{Numerical Scheme}\label{sec:numerical_scheme}
The radiation MHD equations are solved time-explicitly by ZEUS using the
{\color{\oldminor}Method of Characteristics--Constrained
Transport (}MoC--CT{\color{\oldminor})}
algorithm except for the radiation-gas energy exchange terms
$\pm(4\pi B - cE){\kappa}_\text{P}\rho$ and the radiative diffusion term
$-\nabla\cdot\bm{F}$ whose time scales are much shorter than the MHD time
scale \citep{Stone_92,Stone_92b,Turner_01}. Those terms are coupled and solved time-implicitly using
Newton-Raphson iteration and the multi-grid method with a Gauss-Seidel smoother \citep{Tomida_13}.
The gas temperature $T$ used in evaluating the mean 
opacities is fixed to that in the previous time step to linearize the radiative diffusion equation. 

We assume no explicit resistivity and viscosity in the basic equations, and thus
turbulent dissipation occurs through the sub-grid numerical
dissipation. The numerically dissipated energy is captured in
the form of additional internal energy in the gas, {\color{\oldmajor}effectively
resulting in an additional term $Q_\text{diss}$ in the gas energy
equation (\ref{eq:energy_gas})}.    To accomplish this, the original ZEUS algorithm is
modified so as to conserve total energy \citep{Turner_03,Hirose_06}.

During {\color{\oldminor}the} simulations, we employ a density floor of $10^{-6}$ of the initial
midplane density to avoid very small time steps. We also employ a small internal
energy floor for numerical stability \citep{Hirose_06}. 
The total artificial energy injection rate associated with these floors and numerical errors 
in the implicit solver is {\color{\oldminor}generally less than} $\sim 1$ \% of the cooling/heating rates of the final steady state.

\subsection{Initial Conditions}\label{sec:initial_condition}
The initial conditions within the disk photosphere are determined as follows:
First, the vertical profiles of mass density $\rho(z)$, pressure $p(z)$ and temperature $T(z)$, as well
as the initial surface density $\Sigma_0$ and the photosphere height $h_0$, are determined
from an $\alpha$ model described in the Appendix~\ref{sec:dim}, 
by choosing an alpha $\alpha_0$ and an effective temperature ${T_\text{eff}}_0$. 
Then, the internal energy density is calculated
via the EOS $e(z) = e(p(z),T(z))$, and the radiation energy density is
calculated by $E(z) = 4\sigma_\text{B}T^4(z)/c$, assuming that the gas and radiation
are in thermal equilibrium with each other. Above the disk photosphere
$z = h_0$, we assume a uniform, low-density atmosphere with $\rho(z > h_0) =
\rho(0)\times10^{-6}$, $e(z > h_0) = e(h_0)\times10^{-6}$, and $E(z > h_0) =
E(h_0)\times10^{-6}$.

The initial velocity field is the linearized Keplerian shear flow $\bm{v} =
(0,-3/2\Omega x,0)$, whose $x$ and $z$ components are perturbed by random noise
of 0.5 \% of the local sound velocity. 
The initial configuration of the
magnetic field is a twisted azimuthal flux tube (with net azimuthal flux, but
zero net vertical flux), which is placed at the center of the simulation box and
is confined within the photospheres. 
(Note that the volume-integrated vertical magnetic flux is conserved to be zero during the simulation while
the volume-integrated azimuthal magnetic flux is not conserved since it can escape from the box through the vertical boundaries.)
The field strength in the tube is uniform
and the ratio of the poloidal field to the total field is 0.25 at maximum. The
field strength is specified by the parameter $\beta_0$, the ratio of thermal
pressure to magnetic pressure at the center of the flux tube.
{\color{\oldminor}The r}esults do not
depend on $\beta_0$ or the shape of the flux tube so long as the initial
development of the MRI is well resolved.

\subsection{Parameters}
Parameters of the simulations are listed in Table \ref{table}. As described
above, 
the initial conditions are specified by two parameters, surface density
$\Sigma_0$ (or equivalently $\alpha_0$) and effective temperature
${T_\text{eff}}_0$. The table also lists time-averaged surface density $\bar{\Sigma}$ ,
effective temperature $\bar{T}_\text{eff}$, and $\alpha$ in the final
steady state (see Section \ref{sec:results} for the definitions of these quantities).

There are also two numerical parameters: the box size and the number of
cells.  Because we consider dependence on the surface density in this work, it is
desirable that the surface density does not change much during the simulation;
therefore we adjust the box height so that the mass loss rate through the
top/bottom boundaries is less than about 10 \% of the initial mass per
hundred orbits, while the MRI is kept resolved {\color{\minor}reasonably} near the disk midplane. Such a box
is typically $\sim 10$ scale heights tall when measured in terms of the final steady-state
pressure scale height $h_\text{p}$ (see Table \ref{table}).
In the fiducial runs, which are the ones discussed in Section \ref{sec:results}, the numbers of
cells are $(32,64,256)$ and the aspect ratio of the box is
$1:4:8$ in the $x$, $y$, and $z$ directions, respectively. See the table for the absolute size of the box.
To check numerical
convergence, we have also run a wide box version (the box size in both the $x$
and $y$ directions is doubled) and a high resolution version (the resolution is
1.5 times higher and the box length in each direction is 1.2 times larger) for
some selected fiducial runs, which we discuss in Section \ref{sec:robustness}.

\section{Results}\label{sec:results}
The diagnostics of the simulations below are based on
horizontally-averaged vertical profiles, which are recorded every 0.01
orbits. The vertical profile of quantity $f$, for example, is computed as
\begin{align}
\left<f\right>(z,t) \equiv \dfrac{\int\!\!\int f(x,y,z,t) dxdy}{\int\!\!\int dxdy},
\end{align}
where the integrations are done over the full extent of the box in $x$ and
$y$. 

\subsection{Thermal Equilibrium Curve}
One way to characterize the results is in terms of the effective
temperature at the disk photosphere.  Fig. \ref{fig:s-curve} shows this quantity as a function
of surface density for equilibrium states that last over $100$ orbits.\footnote{\color{\oldminor}The solution at the right edge of the lower branch (ws0466F) is an exception; see Section 4.1 for details.} 
The time-averaged effective temperature and surface density are computed as
\begin{align}
& {\color{\oldminor}\bar{T}_\text{eff} \equiv \left[\left(\dfrac{1}{2\sigma_\text{B}}\int\left<Q^-\right>dz\right)^{\frac14}\right]}, \\
& \bar{\Sigma} \equiv \left[\int\left<\rho\right>dz\right],
\end{align}
{\color{\oldminor}
where $\left<Q^-\right>$ is the total cooling rate,
\begin{align}
& \left<Q^-\right> \equiv \dfrac{d}{dz}\left<F_z\right> + \dfrac{d}{dz}\left<(e + E)v_z\right>.
\end{align}}
The brackets $\left[\;\right]$ here denote time averaging over a selected
period of 100 orbits in which the disk is in quasi-steady state and the MRI in
the disk is fairly resolved {\color{\oldmajor}(the period of time averaging in each run is listed in Table \ref{table})},
and the space integration is done for the full extent of the box in
$z$.
As mentioned in the previous section, the surface density can vary due to the mass loss through the 
vertical boundaries, and thus the time averaged surface density $\bar{\Sigma}$ is typically smaller than the initial 
surface density $\Sigma_0$ by only a few percent, rising to at most 9 \% (see Table \ref{table}).


The DIMs based on the $\alpha$ prescription of the stress
generally produce S-shaped thermal equilibrium
curves in this plane, as illustrated by the gray curves. Our simulations show
that ``S-curves'' can also arise as a result of MRI turbulence at
temperatures near the hydrogen ionization transition {\color{\oldminor}as reported by \citet{Latter_12}}.
There are two major solution branches: the upper hot branch ($\bar{T}_\text{eff}
\gtrsim 8000$ K, $\bar{\Sigma} \gtrsim 100$ g cm$^{-2}$) and the lower cool branch
($\bar{T}_\text{eff} \lesssim 3000$ K, $\bar{\Sigma} \lesssim 300$ g cm$^{-2}$).
    {\color{\oldminor}For a limited range of surface density ($100 \lesssim \bar{\Sigma} \lesssim 300$ g cm$^{-2}$),
      {\color{\major}there exist two different stable states for a single value of surface density, showing bistability.}
    }
The disk
is almost fully ionized and optically thick (with total 
optical depth $\tau_\text{tot} > 10^4$) on the upper branch, but it is almost wholly
neutral and {\color{\oldminor}much less} optically {\color{\oldminor}thick} (${\color{\oldminor}2<}\tau_\text{tot} < {\color{\oldminor}14}$)
on the lower branch. (See Table \ref{table} for the value of $\tau_\text{tot}$ in each run.)
These features of the upper and lower branches are consistent with the DIMs. 

The detailed structures of gas temperature as well as ionization fraction are given
in Fig. \ref{fig:temp} for three selected runs: ws0429F, ws0446F, and ws0465F. 
(Radiation temperature is almost identical to gas temperature in these cases and thus is not drawn.)
The temperature is above $10^4$ K and the ionization fraction is
almost unity on the
extreme right end of the upper branch (Fig. \ref{fig:temp}A).
However, as the surface density decreases toward the left edge of the upper branch,
the temperature and ionization fraction can fall from values in this
range near the midplane down to $\sim6000$ K and $\sim10^{-2}$,
respectively, in the atmosphere (Fig. \ref{fig:temp}B). 
{\color{\oldmajor}We expect that this low ionization in the atmosphere will not significantly affect 
MRI turbulence near the midplane, where the gas is fully
ionized.\footnote{\color{\oldmajor}External non-thermal ionization by, for example, soft X-ray irradiation from the
disk-star boundary layer could revive ideal MHD in the atmosphere.}} 
On the lower branch, the temperature is
typically below $2000$ K and the ionization fraction drops to $<10^{-7}$ (Fig. \ref{fig:temp}C).  
Here, the vertical temperature gradient is very shallow due to {\color{\oldminor}the}
low optical depth.

\subsection{Enhancement of $\alpha$}
A new finding here is that $\alpha$ is not
constant, as shown by the colors in Fig. \ref{fig:s-curve}. 
To be precise, we define $\alpha$ as
\begin{align}
\alpha \equiv \dfrac{\left[W_{xy}\right]}{\left[P_\text{thermal}\right]},
\end{align}
where vertically-integrated total stress ${W}_{xy}$ and thermal pressure ${P}_\text{thermal}$ are defined as
\begin{align}
&{W}_{xy}(t) \equiv \int\left<w_{xy}\right>dz, \\
&{P}_\text{thermal}(t) \equiv \int\left<p_\text{thermal}\right>dz .
\end{align}
Here $p_\text{thermal} \equiv p + E/3$ and $w_{xy} \equiv -B_xB_y +
\rho v_x\delta v_y$, where $\delta v_y \equiv v_y + (3/2)\Omega x$. (We include radiation pressure
$E/3$ in the thermal pressure, but its fraction is at most $0.057$
in the largest surface density case on the upper branch.)
Changing the definition of $\alpha$ to a time-average of the
instantaneous ratio, $\left[W_{xy}/P_\text{thermal}\right]$, changes the values by typically a few percent,
and always less than 10 \% for those simulations achieving a quasi-steady state.

Throughout the lower branch and at the right end of the upper
branch, $\alpha$ is approximately $0.03$, typical of values found previously in local numerical
simulations of MRI turbulence that lack net vertical magnetic flux. 
However,
near the left edge of the upper branch ($7000 \lesssim T_\text{eff} \lesssim 10000$ K), $\alpha$ rises
to as much as $0.12$, increasing as the surface density decreases.
This behavior
is in contrast with the DIMs, where constant values of $\alpha$ are assumed on each
branch: $\sim 0.1$ on the upper branch and $\sim 0.01$ on the lower branch (see {\color{\oldminor}Section \ref{sec:dimcomparison}}).
{\color{\minor}As shown below, vertical profiles of the energy transport reveal that the notable change in $\alpha$ in our simulations is associated with thermal convection.}

{\color{\oldminor}In Fig. \ref{fig:cooling}, time-averaged profiles of 
  {\color{\major}radiative heat flux $\bar{F}^-_\text{rad}(z)$, advective heat flux $\bar{F}^-_\text{adv}(z)$, and cumulative heating rate $\bar{F}^+_\text{heat}(z)$ are shown for the cases treated in Fig. \ref{fig:temp}. Here, the heat fluxes are defined as
\begin{align}
& \bar{F}^-_\text{rad}(z) \equiv \left[\left<F_z\right>\right],\\
& \bar{F}^-_\text{adv}(z) \equiv \left[\left<(e + E)v_z\right>\right],\\
& \bar{F}^+_\text{heat}(z)\equiv \int_0^z\left[\left<Q_\text{diss}\right> + \left<-\left(\nabla\cdot\bm{v}\right)p\right> + \left<-\nabla\bm{v}:\mathsf{P}\right>\right]dz.
\end{align}}
From the energy equations (\ref{eq:energy_gas}) and
(\ref{eq:energy_rad}), the thermal energy balance in a steady state is written as
\begin{align}
Q_\text{diss} - \left(\nabla\cdot\bm{v}\right)p -
\nabla\bm{v}:\mathsf{P} = \nabla\cdot\bm{F} +\nabla\cdot\left((e + E)\bm{v}\right),
\label{eq:balance}
\end{align}
where the left hand side is the heating rate (turbulent dissipation and compressional heating)
while the right hand side is the cooling rate (radiative diffusion and advection).}
{\color{\major} Therefore, it is expected in Fig. \ref{fig:cooling} that
\begin{align}
\bar{F}^+_\text{heat}(z) = \bar{F}^-_\text{rad}(z) + \bar{F}^-_\text{adv}(z),
\label{eq:intbalance}
\end{align}
which is a vertically-integrated form of equation (\ref{eq:balance}). 
Actually, the equation (\ref{eq:intbalance}) roughly holds in panel B, and almost exactly holds in panels A and C.
(Note that the gray curve ($\bar{F}^+_\text{heat}$) almost matches the red curve ($\bar{F}^-_\text{rad}$) in panels A and C.)
}

When radiative diffusion
carries the dissipated turbulent energy to the disk surface (Figs. \ref{fig:cooling}A and \ref{fig:cooling}C),
the value of $\alpha$ is typical of MRI turbulence. 
{\color{\major}
On the other hand, when advection plays a major role in the vertical heat transport near the disk midplane
(Fig. \ref{fig:cooling}B), $\alpha$ is large. As shown in the
next subsection, this advective heat transport is associated
with hydrodynamic thermal convection, 
and it is triggered by high opacity that suppresses heat transport by radiative diffusion. We therefore refer to this advective
heat transport as {\it convection} from now on.
}
The energy transported upward by convection is deposited at higher altitude, but below the photosphere, and is then carried by radiative diffusion toward the disk surface.

{\color{\oldminor}A}s shown in Figs. \ref{fig:correlation}A and \ref{fig:correlation}B, the left edge of
the upper branch, where $\alpha$ is large, is exactly the place where convection
carries much of the heat flux. Here, to measure the convection strength, we {\color{\oldminor}define} the {\color{\major}advective} 
fraction as
{\color{\oldmajor}
  the ratio of the advective energy flux to the total energy flux,
\begin{align}
f_\text{\conv} \equiv \dfrac{\int\left[\left<\left(e+E\right)v_z\right>\right]\sgn(z)\left[\left<p_\text{thermal}\right>\right]dz}{\int\left[\left<\left(e+E\right)v_z\right> + \left<F_z\right>\right]\sgn(z)\left[\left<p_\text{thermal}\right>\right]dz},
\end{align}
}
{\color{\oldminor}where the time-averaged pressure $\left[\left<p_\text{thermal}\right>\right]$ is
  used as a weight function to emphasize the regions within the disk photospheres.}\footnote{\color{\oldminor}The {\color{\major}advective} 
  fraction can be negative; when it is,
energy is transported toward the midplane by advection and deposited there.}
The figures also show that $\alpha$ has a strong
correlation with the {\color{\major}advective} 
fraction, which increases as the surface density (and
temperature) decrease on the upper branch. The reason for this trend of {\color{\major}advective} 
fraction
is that much of the opacity on the upper branch
is due to free-free absorption (i.e., Kramers opacity) and rises rapidly with falling
temperature ($\propto T^{-7/2}$){\color{\major}, driving stronger convection}.
%

The reason why $\alpha$ is enhanced when thermal convection exists can be
understood as follows. Convection drives vertical gas motions, which can create
more vertical magnetic field than in the usual MRI turbulence. This assertion is supported
by Fig. \ref{fig:correlation}C, which shows that MRI turbulence is modified when convection exists
so as to increase the time-averaged ratios of the vertical to the radial components for both
velocity and magnetic field. Here the ratios are computed as
\begin{align}
&f_\text{mag} \equiv \dfrac{\int\left[\left<B_z^2\right>\right]\left[\left<p_\text{thermal}\right>\right]dz}{\int\left[\left<B_x^2\right>\right]\left[\left<p_\text{thermal}\right>\right]dz},\\
&f_\text{vel} \equiv \dfrac{\int\left[\left<v_z^2\right>\right]\left[\left<p_\text{thermal}\right>\right]dz}{\int\left[\left<v_x^2\right>\right]\left[\left<p_\text{thermal}\right>\right]dz}.
\end{align}

The strengthened vertical magnetic field enhances
the magnetic stresses since it is the seed for the most powerful, axisymmetric,
modes of the MRI. Indeed, when convection acts, time-averaged and
vertically-integrated stress $\left[W_{xy}\right]$ and the squared vertical magnetic field $\left[\int\left<B_z^2\right>dz\right]$
scale in proportion to one another, both deviating
upwards from the trend ($\propto \bar{\Sigma}^{4/3}$) seen in the non-convective cases
(Fig. \ref{fig:correlation}D). 
(The ratio of Reynolds stress to Maxwell stress hardly changes with $\bar{\Sigma}$ and is always $\sim 0.23$.)
{\color{\major}In contrast, the integrated pressure $\left[P_\text{thermal}\right]$ has a slightly decreasing trend when convection is strong. Apparently, the cooling due to convection lowers the time-averaged equilibrium pressure from the standard trend line despite the increased magnetic stress.}
Therefore, $\alpha$, the ratio of stress to pressure,
increases when convection is stronger.

\subsection{Evidence for True Thermal Convection}\label{sec:evidence}
{\color{\major}In Fig. \ref{fig:cooling}, we also plot the profiles of the squared hydrodynamic Brunt-V\"{a}is\"{a}l\"{a} frequency,}
\begin{align}
&\frac{N^2}{\Omega^2} \equiv \frac{1}{\left[\left<\Gamma_1\right>\right]}\frac{d\ln\left[\left<p\right>\right]}{d\ln z} - \frac{d\ln\left[\left<\rho\right>\right]}{d\ln z},
\label{eq:brunt}
\end{align}
where $\Gamma_1\equiv(\partial\ln p/\partial\ln\rho)_s$ is the generalized adiabatic index {\color{\minor} and $s$ is the {\color{\add}specific} entropy}.
{\color{\major}Note that this hydrodynamic expression of buoyancy frequency is only valid where thermal pressure 
dominates magnetic pressure, i.e. the plasma $\beta$ is larger
than unity, which is interior to the two vertical dotted lines
shown} in the figure.

{\color{\major}
In solution (B), $N^2$ is negative in the midplane regions, precisely where the advective heat transport {\color{\major}$\bar{F}^-_\text{adv}(z)$} is
substantial, indicating that the advective heat transport
in the upper branch discussed in the previous subsection is really thermal
convection associated with unstable hydrodynamical modes.}
We provide additional direct evidence that this is true thermal convection in the next subsection.
{\color{\major}Note that {\color{\add} the large negative values of $N^2$ indicate that} convection does not cause the time-averaged
temperature gradient to be close to the adiabatic value, as supposedly happens in convection zones
in stars.
{\color{\add}In fact, we have measured Mach numbers in the convective velocities as high as 0.1 in this simulation.}
Such superadiabatic gradients are also observed in MRI simulations
with vertical convection by \citet{Bodo_13}.}

Apart from the {\color{\add}case} just mentioned, $N^2$ is generally
positive, indicating convective stability, {\color{\major}provided that $\beta > 1$}.
On the other hand, 
{\color{\add}in solution (B), $N^2$ is negative in some regions where $\beta < 1$,} 
which, however, does not mean that they are convectively
unstable. Because magnetic pressure supports the plasma rather
than thermal pressure alone, the hydrodynamic Brunt-V\"{a}is\"{a}l\"{a}
frequency is not the relevant quantity for
buoyancy instabilities there. Rather,
 {\color{\oldmajor}Parker instabilities can in general act in these regions \citep{Blaes_07}.}

Two solutions near the right edge of the lower branch
(ws0438F and ws0466F)  also show a nonzero advective 
  fraction (Fig. \ref{fig:correlation}B), but do not have an enhanced $\alpha$ (Fig. \ref{fig:correlation}A). 
We have confirmed that $N^2$ is {\color{\add}negative where advective heat transport is
observed in these simulations.} 
{\color{\add}However, the convection in these solutions is too weak to
  strengthen the vertical magnetic field, with convective Mach numbers that are two orders
of magnitude smaller than on the upper branch.}

\subsection{\color{\oldmajor}Convective/Radiative Limit Cycle}
So far we have been discussing trends in the time-averages of the
simulations. However, in a given simulation, convection is often intermittent,
and the system traverses limit-cycles, switching convection on and off episodically.
These limit-cycles are traversed on roughly a thermal timescale.\footnote{\color{\oldminor}The thermal time for each run, computed as $t_\text{th} \equiv \int\left<e+E\right>dz / \int\left<Q^+\right>dz$, is listed in Table \ref{table}.}   In contrast,
the dwarf-nova limit cycles, because they require changes in the local surface
density, are a separate phenomenon and take much longer, of order an inflow timescale.

{\color{\oldminor}An example of a convective/radiative limit cycle is seen} in Fig. \ref{fig:evolution}, where time variations of the vertically-integrated pressure $\tilde{P}_\text{thermal}(t){\color{\oldminor} \equiv \left\{P_\text{thermal}\right\}}$,
stress $\tilde{W}_{xy}(t) {\color{\oldminor}\equiv \left\{W_{xy}\right\}}$
and instantaneous {\color{\major}advective} 
fraction $\tilde{f}_\text{\conv}(t)$ {\color{\oldminor}defined as
  {\color{\oldmajor}
\begin{align}
\tilde{f}_\text{\conv}(t) \equiv \left\{\dfrac{\int\left\{\left<\left(e+E\right)v_z\right>\right\}\sgn(z)\left<p_\text{thermal}\right>dz}{\int\left\{\left<\left(e+E\right)v_z+F_z\right>\right\}\sgn(z)\left<p_\text{thermal}\right>dz}\right\}
\end{align}
}
are shown for $\Sigma_0 = 140$ g cm$^{-2}$ on the upper branch (ws0441F).
Here the brackets $\left\{\;\right\}$ denote box-car smoothing over a width of one orbit.
{\color{\oldminor}Also shown are}
  the instantaneous $\alpha$ and ratios of the vertical to the radial components for velocity and magnetic fields defined as
\begin{align}
&{\color{\oldminor}\tilde{\alpha}(t) \equiv \dfrac{\tilde{W}_{xy}(t)}{\tilde{P}_\text{thermal}(t)}},\\
&\tilde{f}_\text{mag}(t) \equiv \left\{\dfrac{\int\left<B_z^2\right>\left<p_\text{thermal}\right>dz}{\int\left<B_x^2\right>\left<p_\text{thermal}\right>dz}\right\},\\
&\tilde{f}_\text{vel}(t) \equiv \left\{\dfrac{\int\left<v_z^2\right>\left<p_\text{thermal}\right>dz}{\int\left<v_x^2\right>\left<p_\text{thermal}\right>dz}\right\}.
\end{align}
The curve of {\color{\major}advective} 
fraction demonstrates that convection occurs episodically, anti-correlated with the variation of pressure, indicating that convection is controlled by the temperature-sensitive opacity. The figure also shows that the ratios $\tilde{f}_\text{mag}(t)$ and $\tilde{f}_\text{vel}(t)$, whose time-averaged versions are enhanced when convection acts as discussed above, are actually enhanced at the beginning of each of the convective episodes. 
We interpret this as being due to the generation of vertical magnetic field by the onset of vertical convection, which seeds the axisymmetric MRI.   The vertical to radial magnetic field ratio then falls back to the usual value as horizontal field is built by the
MHD turbulence.
{\color{\oldminor}The figure also shows that the stress begins to increase when the convection is fully developed and is followed by pressure with a finite time lag of several orbits. The stress parameter $\alpha$, which is already higher than that of normal MRI turbulence, is further amplified when stress is high while pressure is low.}

{\color{\oldmajor}
That thermal convection is actually {\color{\oldmajor} operating during what
we are calling the convective periods} ($\tilde{f}_\text{\conv} \sim 1$) is clearly demonstrated by Fig. \ref{fig:slice_convective},
in which various quantities on a selected $x$-$z$ plane ($y = 0$) at $t = 103$ orbits are shown.
The {\color{\add}specific} entropy\footnote{\color{\add}Specific entropy is computed as a function of $(\rho, e/\rho)$ like other thermodynamical variables. See equation (A54) in \citet{Tomida_13}.} is highest near the midplane, 
which drives low-density and high-temperature plumes that coherently transport
heat upward (i.e. $ev_z$ is mostly positive (negative) for $z > 0$ ($z < 0$), respectively).
We therefore expect coherent vertical magnetic fields will be generated on the scale of the convective plumes, which 
is about half the pressure scale height.\footnote{\color{\oldminor}See Table \ref{table} for the pressure scale height in each run.}
Note that strong isolated magnetic fluxes are also associated with low density blobs, which suggests that the finite amplitude, slow mode buoyancy mechanism 
that we pointed out in \citet{Blaes_11} also contributes to vertical advection
of heat. However, it is now completely dominated by the genuine thermal convection that fills much
of the volume. These features in a convective period are contrasted with those in a radiative period at $t = 90$ orbits  (Fig. \ref{fig:slice_radiative}). 
The vertical gradient of {\color{\add}specific} entropy is now {\color{\add}almost zero}, indicating that the disk is convectively {\color{\add}neutrally} stable. {\color{\add}This adiabatic gradient is caused by convection in the preceding convective period.}
Also, 
we see mostly random motions and only the slow mode mechanism is operating for the small net vertical advection of heat; 
however, the main heat transport mechanism here is, 
of course, radiative diffusion ($\tilde{f}_\text{\conv} \sim 0$). 
}

{\color{\oldminor}It might be instructive to visualize the convective/radiative limit cycles as }trajectories in the pressure vs. stress plane.
Fig. \ref{fig:limitcycle} shows {\color{\oldminor}such} trajectories for $\Sigma_0 =$ 140 (ws0441F), {\color{\oldminor}which we discussed above, as well as $\Sigma_0 = $} 191 (ws0427F), 248 (ws0472F), 402 (ws0469F), and 1075 g cm$^{-2}$ (ws0429F) on the upper branch, in terms of
the time variation of the vertically-integrated pressure $\tilde{P}_\text{thermal}(t)$ and stress $\tilde{W}_{xy}(t)$.
{\color{\oldminor}The range of time is $t_1 < t < t_2$, where $t_1$ and $t_2$ are given in Table \ref{table}.}
The color intensity represents the instantaneous {\color{\major}advective} 
fraction $\tilde{f}_\text{\conv}(t)$.
When {\color{\oldminor} the surface density is higher} and convection is not
present ($\Sigma_0 = 402$ and $1075$ g cm$^{-2}$), {\color{\oldminor}no limit cycle is seen, as} pressure is almost unchanged in
the face of stress fluctuations and thus the trajectories are almost vertical.
{\color{\oldminor}As the surface density decreases ($\Sigma_0 = 248$, $191$  and $140$ g cm$^{-2}$), 
the pressure fluctuation becomes larger and}
a limit cycle running clockwise in the plane is established: 
(1) magnetic stress is strengthened in fully-developed convection,
(2) stronger magnetic turbulence leads to greater dissipation, which increases
the temperature and therefore the pressure; (3) higher temperature reduces the 
opacity, suppressing convection, and (4) without convection, magnetic fields
weaken and the temperature declines, increasing the opacity, which eventually
restores convection.  On the left edge of the trajectories in the lower surface
density (convective) cases, stress increases while pressure stays low; this
phase lag further {\color{\oldminor}increases} $\alpha$ {\color{\oldminor}as we discussed above}.
%

\section{Discussion}\label{sec:discussion}
\subsection{Runaway Heating and Cooling}\label{sec:runaway}

We always find runaway cooling (heating) of the disk
beyond the left (right) edge of the upper (lower) branch, respectively. In
Fig. \ref{fig:edge}, we show time trajectories of such runs in the surface density
vs. effective temperature plane: ws0488R showing runaway cooling and ws0467R
showing runaway heating. The initial development of MRI turbulence
in both runs was similar to that in the other runs. However, both runs passed
by the edge of the nearest stable
branch and did not reach a steady state, which indicates that the two stable
branches are actually truncated there.
{\color{\oldmajor}These facts indicate that any limit cycle in this plane runs
anti-clockwise, which is consistent with the DIMs, where the cooling rate (heating rate)
always exceeds the heating rate (cooling rate) beyond the left (right) edge of
the upper (lower) branch, respectively.
In fact, signs of the thermal runaway for the state transitions are seen near the edges of the branches.
For example, the disk at the right edge of the lower branch (ws0466F) 
stayed in thermal equilibrium for $\sim 80$ orbits and then began to flare up; on the other hand,
the disk at the left edge of the upper branch (wt0442F) 
began to collapse after thermal equilibrium of $\sim 100$ orbits.
{\color{\oldmajor}Similar behavior at the} edges of the two stable branches
{\color{\oldmajor} was reported} by \citet{Latter_12}.}

Run ws0488R was stopped at $t \sim 50$
orbits since the disk collapsed and resolving MRI near the disk midplane badly
failed. Run ws0467R was also stopped at $t \sim 100$ orbits when the mass loss
from the simulation box became substantial ($\sim 25$\%).\footnote{\color{\oldmajor}The trajectory bends to the left due to this substantial mass loss.}
{\color{\oldmajor}The runs near the edges of the branches (ws0466F and wt0442F) were
also stopped for the same reasons.} To further simulate the thermal runaways
or state transitions between the upper and lower branches, we would need to
dynamically change the box size so that the MRI is always well resolved and the
mass loss is kept small enough.

We could not find long-lived equilibria between
the upper and lower branches{\color{\oldmajor}, suggesting that the
  negative sloped portions of the alpha-based S-curves are unstable}.
In fact, in a few fiducial runs we found
equilibria that lasted more than several tens of orbits, but these were
not fully reproduced when the box size or the resolution was changed.


\subsection{Numerical Robustness of the Results}\label{sec:robustness}
To check the robustness of our results, we performed two kinds of tests; one to
check dependence on the initial conditions (Fig. \ref{fig:diff}) and the other to check
numerical convergence (Fig. \ref{fig:conv}). 

For four arbitrarily selected fiducial runs, we ran a supplementary simulation
whose surface density is almost the
same, but whose initial effective temperature is different from the fiducial
run. In Fig. \ref{fig:diff}, the initial and final states of the four pairs of fiducial and
supplementary runs (ws0441F and ws0494C, ws0464F and ws0465C, ws0471F and
ws0472C, and ws0468F and ws0469C) are shown. The final states of paired runs are
roughly the same both in effective temperature and $\alpha$, which confirms that
the two branches are actually unique attractors. 

We also arbitrarily selected five fiducial runs on the upper branch, $\Sigma_0=1075$ (ws0429F),
$402$ (ws0469F), $247$ (ws0471F), 179 (ws0437F), and 132 g cm$^{-2}$ (ws0446F), and for each we ran two supplementary
simulations: doubling the horizontal box size in one, and increasing the
resolution by 1.5 times and the box size by 1.2 times in the other one. As shown
in Fig. \ref{fig:conv}, the final states of the fiducial run and the corresponding two
supplementary runs are roughly the same, both in effective temperature and $\alpha$
for all five cases. We may therefore conclude that our two main results, the
thermal equilibrium curve and the high $\alpha$ near the edge of the upper branch,
are not sensitive to the box size or the resolution.


{\color{\oldmajor}
\subsection{Comparison with the DIM Model}\label{sec:dimcomparison}
As shown in Fig. \ref{fig:s-curve}, those simulations that do not exhibit
convection lie on our DIM-model curve with fixed $\alpha=0.03$ on the upper
branch, while they are slightly below the curve on the lower branch \citep[cf.][]{Latter_12}. We suspect that the discrepancy may come from our DIM's
assumption that the disk is very optically thick even on the lower branch,
where the measured optical depth in the simulations can be as low as 2.3
(Table 1).

On the other hand, our results near the left edge of the upper branch deviate upwards from the fixed $\alpha$ curve as $\alpha$ increases due to convection. In some cases, however, our results with larger values of $\alpha$ lie below a DIM-model curve computed assuming a smaller $\alpha$. Also, the minimum (maximum) surface density of our upper (lower) branch, respectively, are larger than those of the relevant DIM-model curve. These discrepancies are presumably due to our neglect of convection in computing the DIM curves because convection tends to increase the critical surface densities \citep[see, for example,][]{Pojmanski_86}.
} 

{\color{\oldmajor}
\subsection{Effect of the Initially Imposed Net Toroidal Flux}
It is widely known that the saturation of MRI turbulence in the local shearing box depends on the net vertical flux or net toroidal flux threading it \citep[see, for example,][]{Hawley_95,Latter_12}. Although we do not impose a net vertical flux in the box, we do impose a net toroidal flux initially in the box (see Section \ref{sec:initial_condition}). Therefore one might argue that the saturation of MRI turbulence in our simulations could be affected by the initial net toroidal flux. 

Stratified shearing box simulations, however, generally do not retain a net toroidal flux because of buoyant escape through the vertical boundaries.
Fig. \ref{fig:toroidalflux} shows the time variation of the net toroidal flux $\tilde{\Phi}_y(t)\equiv \left\{\int\left<B_y\right>dz\right\}$ as well as the net radial flux $\tilde{\Phi}_x(t)\equiv \left\{\int\left<B_x\right>dz\right\}$ for the initial hundred orbits for three selected runs: a convective solution on the upper branch (ws0441: $\alpha = 0.0927$), a radiative solution on the upper branch (ws0429: $\alpha = 0.0332$), and a radiative solution on the lower branch (ws0435: $\alpha = 0.0312$).
The net toroidal flux in every case fluctuates significantly in time, flipping
in sign, and there is no indication of any memory of the initial net toroidal
flux.  (The reason for the sign flips is presumably that azimuthal flux arises
from shear of radial flux, which also flips sign over time due to the still
poorly understood dynamo of stratified MRI turbulence.) Therefore, we conclude that the saturation of MRI turbulence in our simulations is independent of the initially imposed toroidal flux.
}

{\color{\oldmajor}
\subsection{Alternative Explanations for Large $\alpha$ in the Outburst Phase}


We have shown that convection enhances MRI turbulent stress, which can
{\color{\oldmajor} increase} $\alpha$ above $0.1$.
Since convection necessarily occurs in the outburst phase due to the strong
temperature dependence of opacity,
we have found an $\alpha$-enhancement mechanism that is due to the internal
physics within the disk in this regime.

{\color{\oldmajor}There are other external mechanisms, however, that might be
considered candidates for producing enhanced $\alpha$.  For example, it
is well-known that imposition of net vertical magnetic flux raises the
saturation level of the turbulence.}
However, it is also known that the dependency of stress
on net vertical field ${B_z}_\text{net}$ is fairly strong{\color{\major}, $\propto {{B_z}_\text{net}}^2$ \citep{Suzuki_10,Okuzumi_11}}.
Therefore, the fact that the observed $\alpha$ in the outburst phase is always of order 0.1 is a strong
constraint on the existence of a global net vertical field
in the disk in the outburst phase \citep{King_07}.  On the other hand, global
simulations of MRI turbulent disks have produced local net vertical fluxes
through magnetic linkages in the disk corona \citep{Sorathia_10}.  It could
be that this mechanism might play a role in producing large $\alpha$'s in the
outburst phase, but why this would only occur in the outburst phase is unclear.
{\color{\oldmajor}\citet{Sorathia_12} have also suggested that the large
$\alpha$'s inferred in the outburst phase may be due to transient periods of
magnetic field growth in the jump to outburst, together with gradients
in the global disk.  For these}
and many other reasons, thermodynamically consistent, global MRI
simulations of disks in the hydrogen ionizing regime will be of great interest.

Another point worth mentioning is that an enhanced stress does not necessarily
lead to an enhanced $\alpha$.   If, for example, pressure also rises in proportion to the
enhanced stress (via enhanced dissipation), $\alpha$, the ratio of stress to pressure, would not be {\color{\oldmajor} increased.
  {\color{\major}In our simulations, the convective cooling controls the pressure well enough to lead to an increase in $\alpha$. Net vertical flux, which increases stress, could in principle also explain the high $\alpha$ in the high state, but whether it does or not will depend on the scaling of cooling rate with pressure in the presence of vertical flux.  New simulations that carefully account for thermodynamics will be necessary to determine this scaling.}
}
}

{\color{\oldmajor}
\subsection{Relation to the Radiation Pressure Dominated Thermal Instability}
We have remarked on the possible thermal instability of the S-curve branch that
should link the low and high states.   Any such instability would be qualitatively
different from any thermal instability that affects a radiation pressure dominated regime
\citep{Shakura_76,Turner_04,Hirose_09,Jiang_13}.
In the temperature range between the low and high states relevant to dwarf
novae, the opacity increases rapidly with increasing temperature because this
is the regime in which H ionizes, so that 
the dominant opacity source is free-free or bound-free absorption of the negative hydrogen ion H$^-$.
Thus, an upward fluctuation in heating
receives positive feedback because it is accompanied by weaker cooling.
By contrast, thermal fluctuations in the radiation-dominated regime are aided
both by the sensitivity of radiation pressure to temperature ($\propto T^4$)
and possible dynamical coupling between total thermal pressure (gas plus
radiation) and heating associated with MHD turbulence \citep{Jiang_13}.

}

\section{Conclusions}\label{sec:conclusion}
We have successfully identified two distinct stable branches of
thermal equilibria in the hydrogen ionization regime of accretion disks:  a
hot ionized branch and a cool neutral branch.  We have measured
high values of $\alpha$ on the upper branch that are comparable to those
inferred from observations of dwarf nova outbursts, the very systems where
$\alpha$ is measured best.  The physical mechanism
for creating these high $\alpha$ values is specific to the physical conditions
of the hydrogen ionization transition that is responsible for these outbursts.
That mechanism is thermal convection triggered by the strong dependence of
opacity upon temperature.
We confirm the finding of \citet{Bodo_12} that convection modifies the
MRI dynamo to enhance magnetic stresses, but our more realistic
treatment of opacity and thermodynamics yields a larger effect, with
a substantial increase in $\alpha$.
Convection acts only in a narrow range of temperatures near the
ionization transition because that is where the opacity is greatest. Thus the
high values of $\alpha$ are restricted to the upper bend in the S-curve.
Because the observational inference of high values of $\alpha$ is based on outburst
light-curves, our finding that $\alpha$ is especially large near the low surface density end of the
upper branch is relevant to the quantitative interpretation of these light-curves.
Similarly, when we understand better the stresses in the plasma on the lower
branch, where non-ideal MHD effects are important \citep[see, for example,][]{Menou_00}, those results will bear on
observational inferences tied to the recurrence times of dwarf novae.






\acknowledgments

 We thank {\color{\add}G. Ogilvie,} J.-P. Lasota, I. Kotko, J. Stone, J. Hawley, S. Inutsuka and N. Turner for
  useful discussions. {\color{\oldminor}We thank the referee for his/her valuable comments and suggestions.} 
  This work was supported by JSPS KAKENHI Grant 24540244 and 23340040,
  joint research project of ILE, Osaka University (SH), NSF Grant AST 0707624(OB), and NSF Grant AST-0908336 and NASA/ATP Grant
  NNX11AF49G (JK). Numerical simulations were carried out on Cray XC30 at CfCA, National Astronomical Observatory of Japan,
  SR16000 at YITP in Kyoto University, and XSEDE systems Stampede and
  Kraken (supported by NSF Grant TG-MCA95C003).






\appendix

\section{Thermal equilibria based on the alpha prescription}\label{sec:dim}
We computed thermal equilibria based on the alpha prescription (Fig. \ref{fig:s-curve})
following Mineshige and Osaki (1983), except that we always assumed optically
thick disks and did not consider convection for simplicity. The basic equations
for pressure $p(z)$ and temperature $T(z)$ describe hydrostatic and radiative
equilibrium in the vertical direction:
\begin{align}
  &\dfrac{dp}{dz} = -\rho(p,T)\Omega^2z, \\
  &\dfrac{dT}{dz} = \dfrac{3\kappa_\text{R}(\rho,T)}{16\sigma_\text{B}T^3}\dfrac{F(z)}{\Omega^2 z}\dfrac{dp}{dz}.
\end{align}
The EOS $\rho = \rho(p,T)$ and Rosseland-mean opacity $\kappa_\text{R}(\rho,T)$
here are the same as those employed in the simulations. The
radiative flux $F(z)$ is given as a function of height $z$ as $F(z) = \sigma
{T_\text{eff}}_0^4\min(1,z/(h_0/2))$ to guarantee physical boundary conditions
$F(0)=0$ and $F(h_0) = \sigma_\text{B} {T_\text{eff}}_0^4$, where $h_0$ is the
photosphere height (defined as the height where optical depth is $2/3$) and
${T_\text{eff}}_0$ is the effective temperature.

We integrate the equations from the disk photosphere $z = h_0$ toward the
midplane $z = 0$ using the boundary condition $T(h_0) = {T_\text{eff}}_0$ and
$p(h_0) = p(\rho_0,{T_\text{eff}}_0)$. Here, the parameters are $h_0$ and
${T_\text{eff}}_0$, and the density at the photosphere $\rho_0$ is 
determined by the condition that the optical depth down to the photosphere is
2/3:  $\kappa_\text{R}(\rho_0,{T_\text{eff}}_0)p(\rho_0,{T_\text{eff}}_0) =
(2/3) \Omega^2 h_0$. 

To obtain a thermal equilibrium curve for a fixed alpha $\alpha_0$, we seek the
photosphere height $h_0$ that satisfies the alpha prescription
$(3/2)\Omega\alpha_0\int_0^{h_0}p(z)dz =
\sigma_\text{B}{T_\text{eff}}_0^4$. Once $h_0$ is found, we have equilibrium
profiles of pressure $p(z)$, temperature $T(z)$, and density $\rho(z) =
\rho(p,T)$ for the specified $\alpha_0$ and effective temperature
${T_\text{eff}}_0$, from which we can compute the corresponding surface density
as $\Sigma_0 = 2\int_0^{h_0}\rho(z)dz$. Repeating this procedure for different
effective temperatures ${T_\text{eff}}_0$ at various fixed values of $\alpha_0$,
we obtain the surface density as a function of effective temperature $\Sigma_0 = \Sigma_0({T_\text{eff}}_0; \alpha_0)$, which is the thermal equilibrium curve for each $\alpha_0$.

\clearpage




\begin{figure}
  \centering
  \includegraphics[scale=0.75]{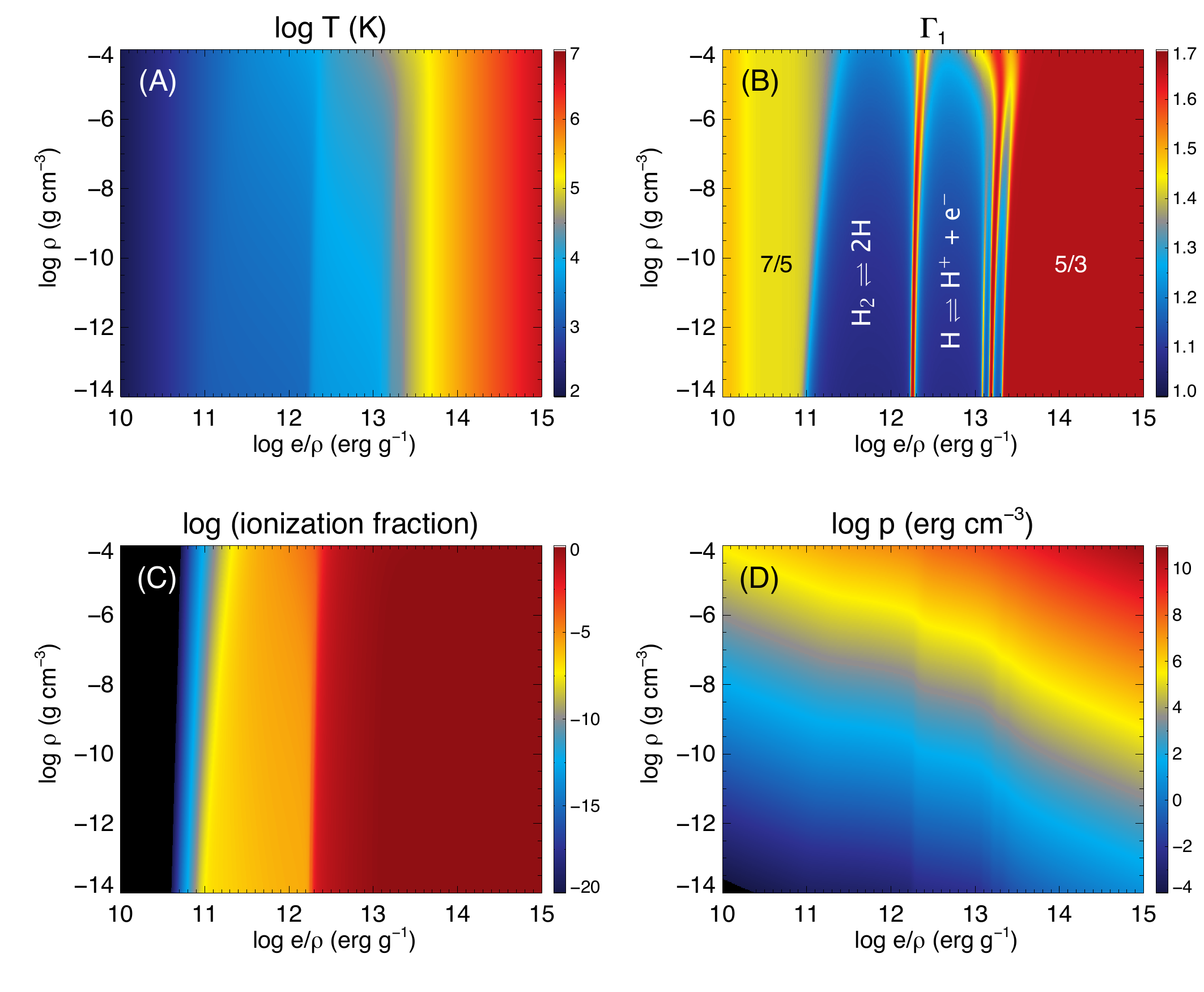}
\caption{Non-ideal EOS computed from Saha equations: Gas
  temperature (A), generalized adiabatic index $\Gamma_1
  \equiv (\partial\ln p/\partial\ln\rho)_s$ (B), ionization
  fraction (C), and pressure (D) as a function of
  specific energy density $e/\rho$ (erg g$^{-1}$) and mass density
  $\rho$ (g cm$^{-3}$).}\label{fig:eos}
\end{figure}

\begin{figure}
  \centering
  \includegraphics[scale=0.75]{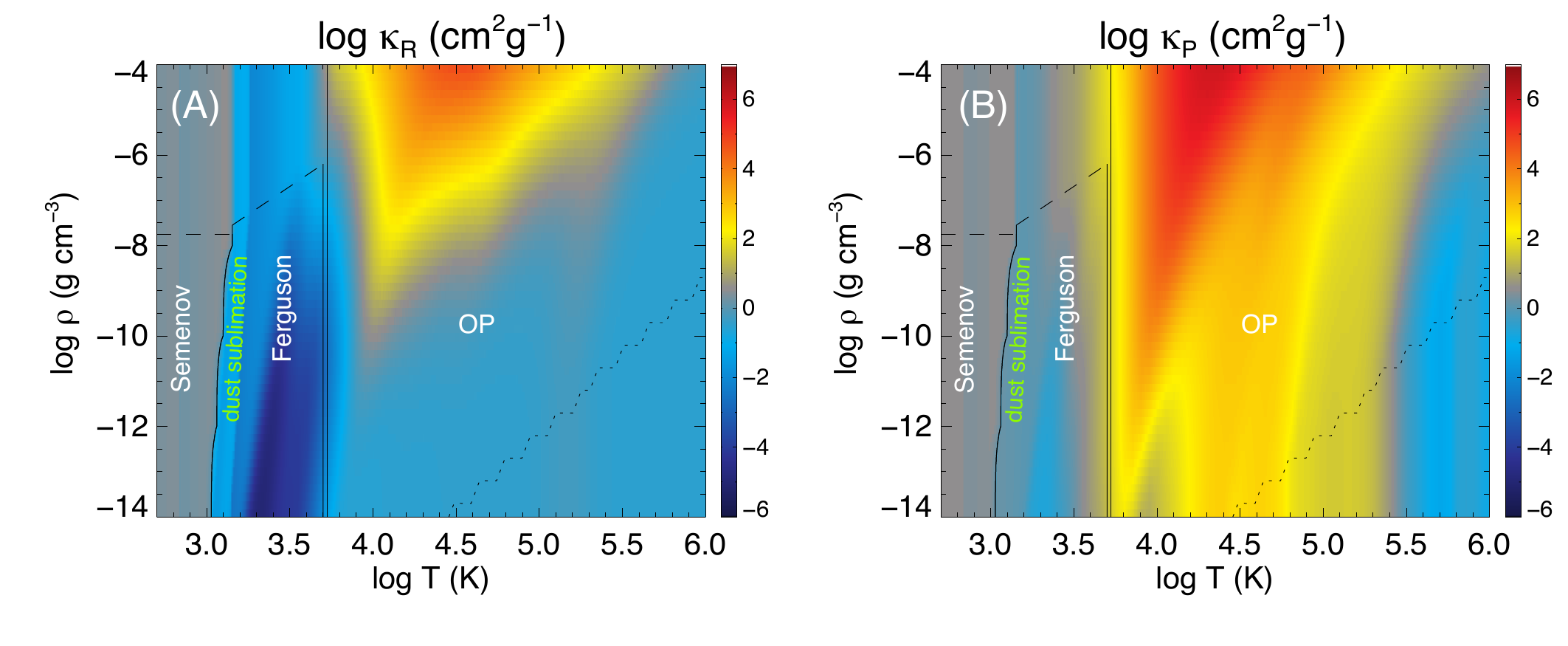}
\caption{Mean opacities combining three published opacity tables by
  Semenov et al. (2003), Ferguson et al. (2005) and the Opacity Project
  (OPCD\_3.3): Rosseland-mean opacity (A) and Planck-mean opacity (B) as
  a function of gas temperature $T$ (K) and mass density $\rho$ (g cm$^{-3}$). 
  The solid curves are boundaries between adjacent opacity tables.
  The dashed curve denotes the
  upper bound of Semenov's and Ferguson's opacities while the dotted curve
  denotes the lower bound of OPCD\_3.3.
}\label{fig:opacity}
\end{figure}

\begin{figure}
  \centering
  \includegraphics[scale=0.75]{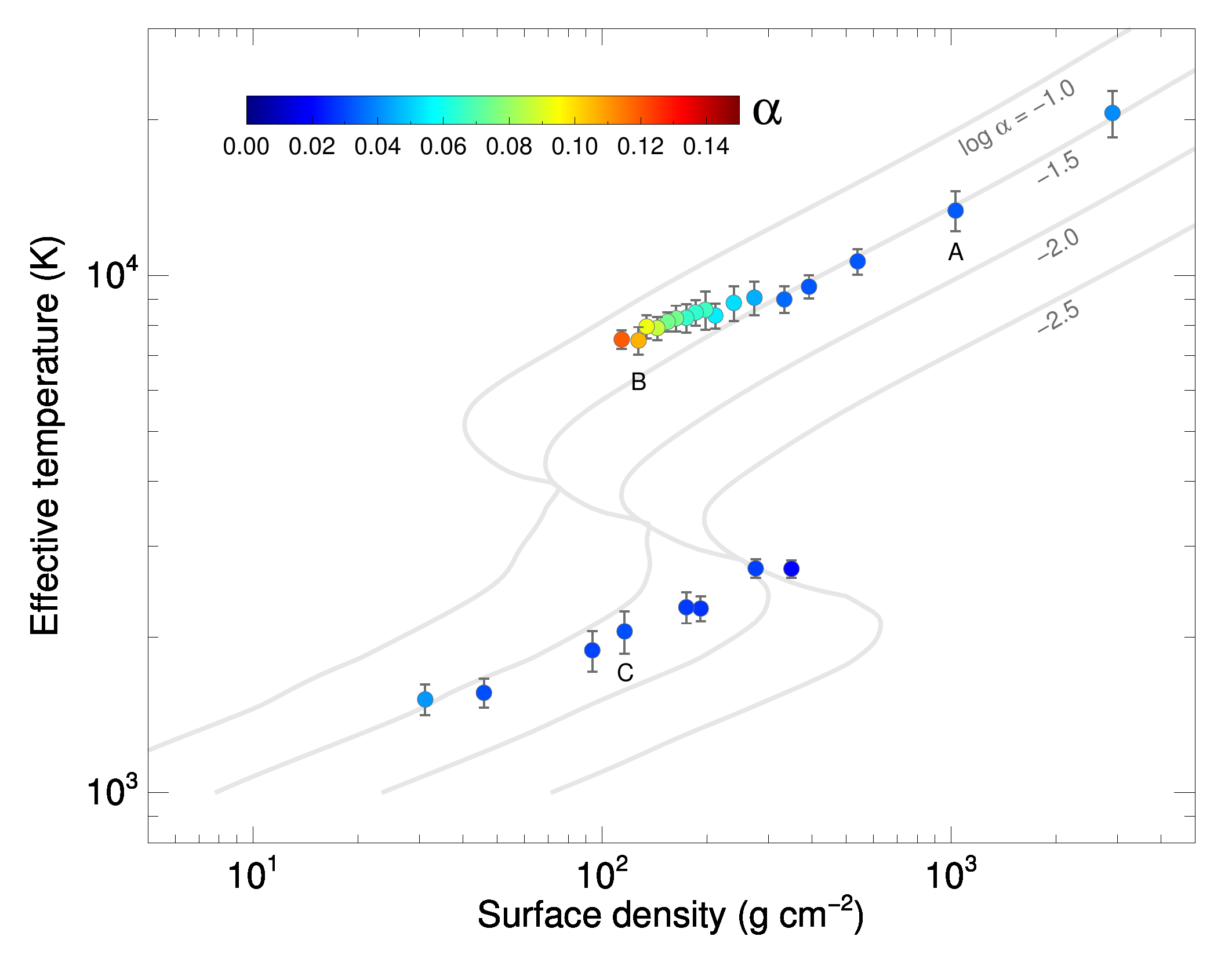}
\caption{Time-averaged effective temperature $\bar{T}_\text{eff}$ vs. surface density $\bar{\Sigma}$ in
  thermal equilibrium states.
{\color{\oldminor}Error bars represent one standard deviation in the time
variability of effective temperature $T_\text{eff}(t)\equiv\left(\int\left<Q^-\right>dz/{2\sigma_\text{B}}\right)^{1/4}$.}
Colors represent the time-averaged
  values of $\alpha$. Gray curves are thermal equilibria produced by a DIM based on
  the alpha prescription.
Solutions labeled with A, B, and C correspond to panels A, B, and C, respectively in Figs. \ref{fig:temp} and \ref{fig:cooling}.}\label{fig:s-curve}
\end{figure}

\begin{figure}
  \centering
  \includegraphics[scale=0.375]{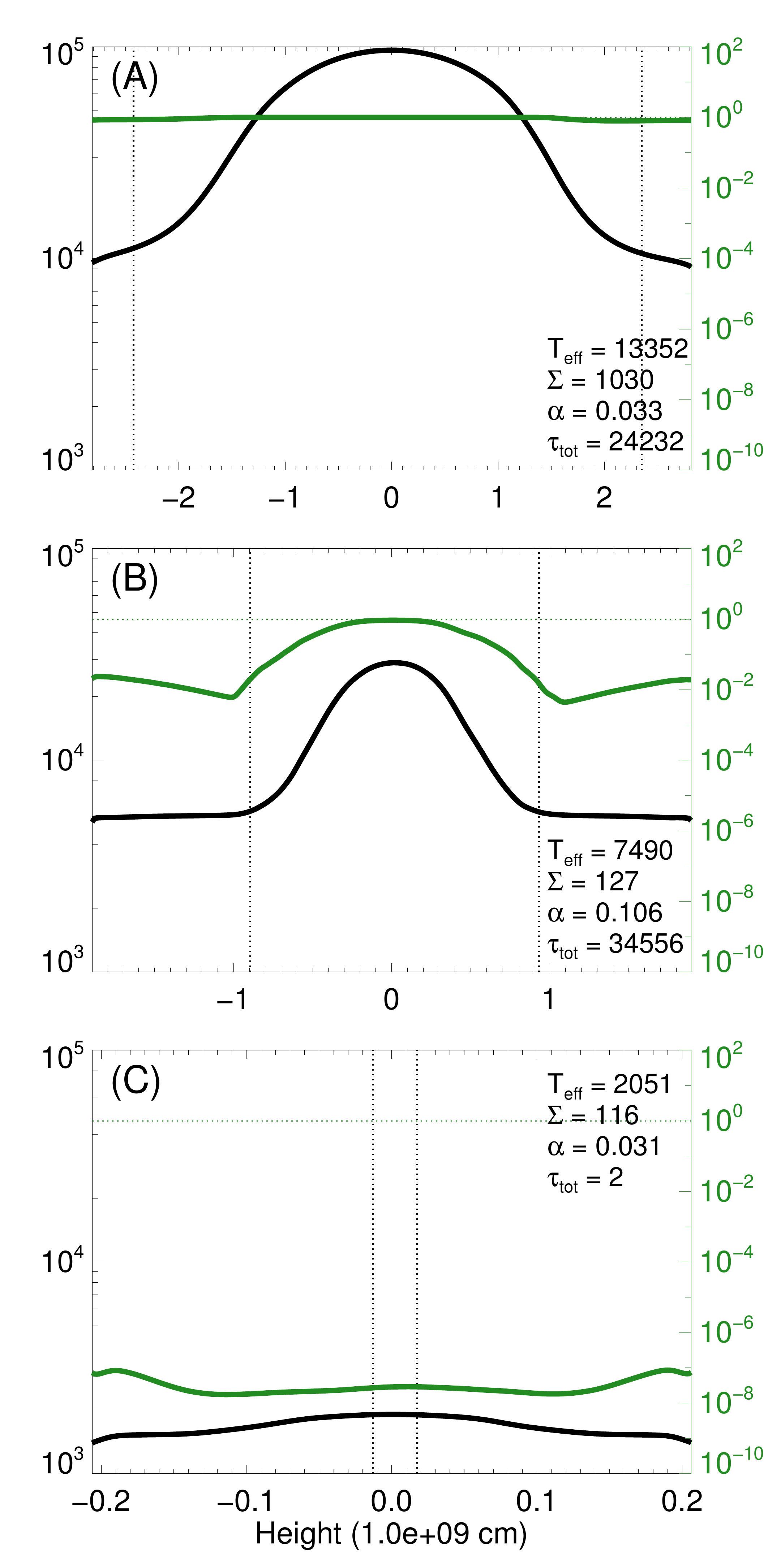}
\caption{Time-averaged vertical profiles of gas temperature (K) (black) and ionization fraction (green) 
  for two upper branch simulations, (A) $\Sigma_0 = 1075$ g cm$^{-2}$ {\color{\oldminor}(ws0429F)} and (B) $\Sigma_0 = 132$ g cm$^{-2}$ {\color{\oldminor}(ws0446F)}, and one lower branch simulation, (C) $\Sigma_0 = 120$ g cm$^{-2}$ {\color{\oldminor}(ws0465F)}. 
{\color{\oldminor}The axis for the green curves is on the right.}
The vertical dotted lines denote the heights where the Rosseland-mean optical depth from the top/bottom boundary is unity.
Also, in each frame, the corresponding values of $\bar{T}_\text{eff}$ (K), $\bar{\Sigma}$ (g cm$^{-3}$), $\alpha$, and the total optical depth $\tau_\text{tot}$ are shown.}\label{fig:temp}
\end{figure}

\begin{figure}
  \centering
  \includegraphics[scale=0.375]{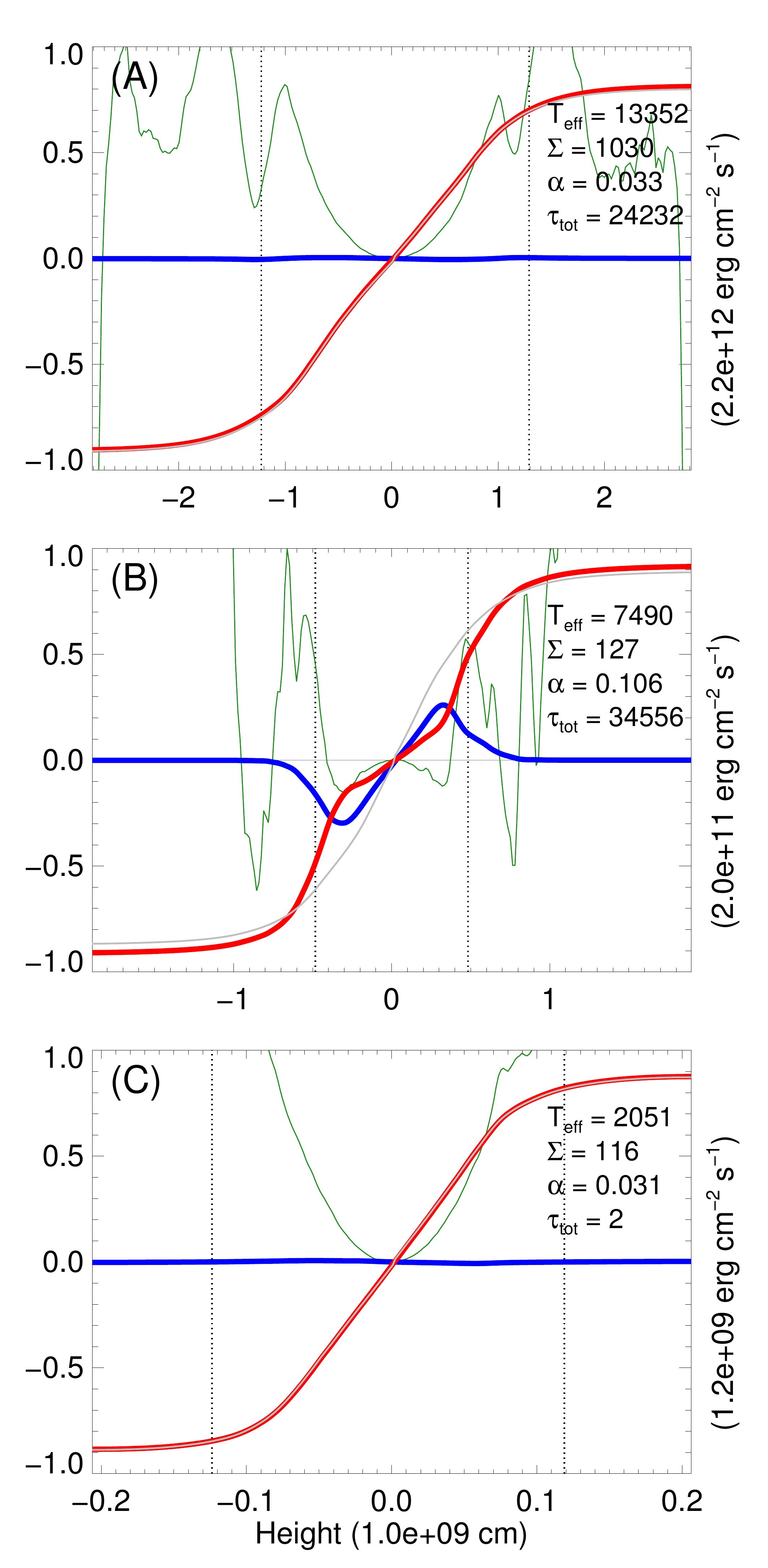}
\caption{Time-averaged vertical profiles of 
{\color{\major}radiative heat flux $\bar{F}^-_\text{rad}$ (red), advective heat flux $\bar{F}^-_\text{adv}$ (blue), and cumulative heating rate $\bar{F}^+_\text{heat}$ (gray), normalized by the value shown on the right axis in each panel,} for the cases shown in Fig. \ref{fig:temp}.
{\color{\oldmajor}The {\color{\major}green curves show} {\color{\add}$N^2/\Omega^2$}, where $N$ is the hydrodynamic Brunt-V\"{a}is\"{a}l\"{a} frequency.} 
{\color{\major}The plasma $\beta$ is larger than unity at the heights between the two vertical dotted lines.}
Other notations are the same as in Fig. \ref{fig:temp}.}\label{fig:cooling}
\end{figure}


\begin{figure}
  \centering
  \includegraphics[scale=0.30]{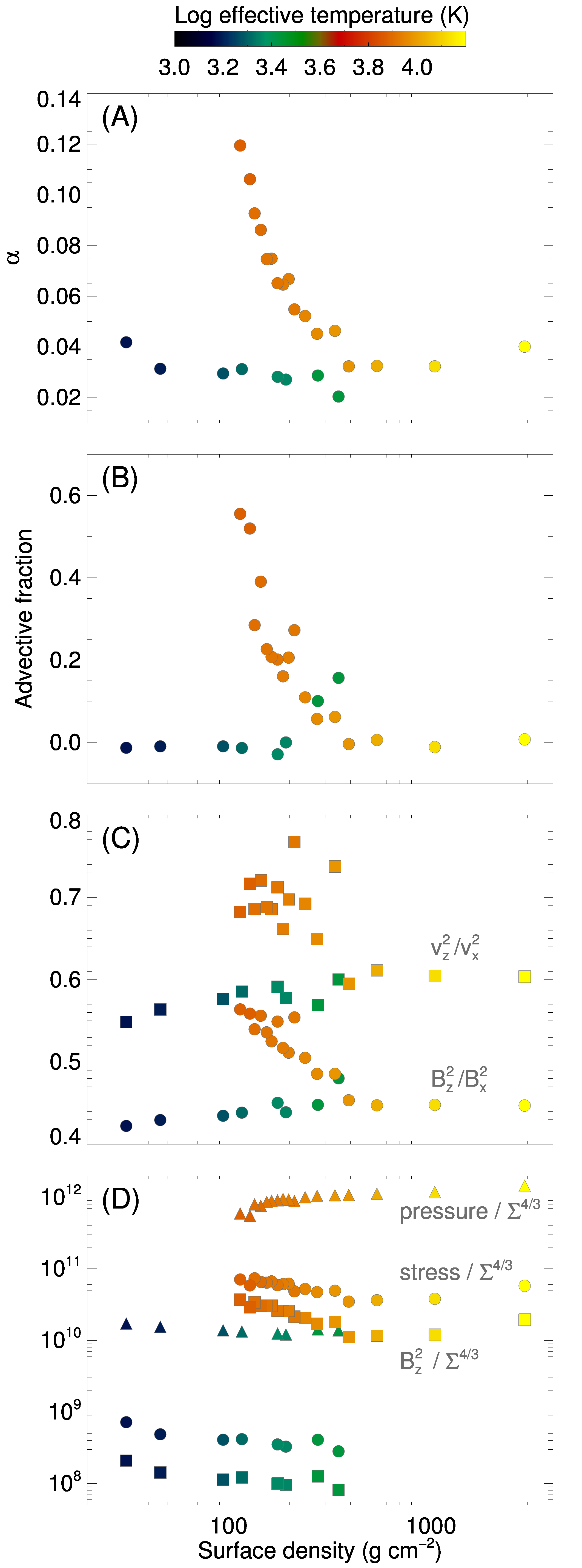}
\caption{Various time-averaged quantities as a function of the surface density $\bar{\Sigma}$:
  (A) $\alpha$, (B) {\color{\major}advective} 
  fraction $f_\text{\conv}$, (C) ratios of the vertical to the radial components for velocity field $f_\text{vel}$ (squares) and for magnetic field $f_\text{mag}$ (circles), (D) vertically-integrated square of vertical magnetic field $\left[\int\left<B_z^2\right>dz\right]$ (squares), total stress $\left[W_{xy}\right]$ (circles), and thermal
  pressure $\left[P_\text{thermal}\right]$ (triangles), each divided by $\bar{\Sigma}^{4/3}$. Colors represent the
  time-averaged effective temperature $\bar{T}_\text{eff}$ in each simulation. The two vertical
  dotted lines indicate the surface density range $100 \lesssim \bar{\Sigma} \lesssim 350$ (g cm$^{-2}$), where convection acts on the upper branch.}\label{fig:correlation}
\end{figure}

\begin{figure}
  \centering
  \includegraphics[scale=0.525]{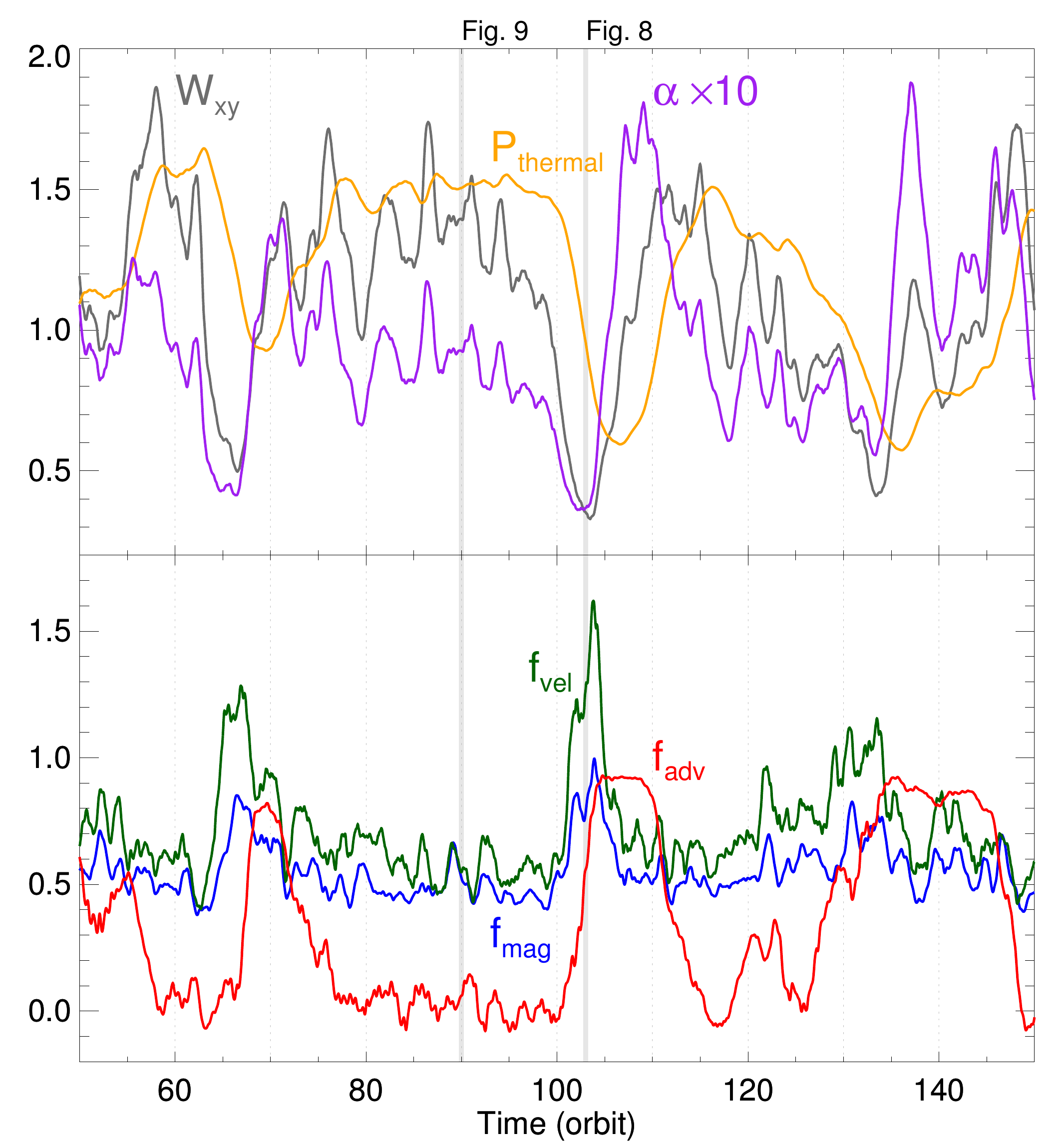}
  \caption{Time variations of various quantities for $\Sigma_0 = 140$ g cm$^{-2}$ on the upper branch {\color{\oldminor}(ws0441F)}. Upper panel: Vertically-integrated total stress $\tilde{W}_{xy}$ (gray) and thermal pressure $\tilde{P}_\text{thermal}$ (orange), and $\tilde{\alpha}\times10$ (purple). The total stress and thermal pressure are normalized arbitrarily here. Lower panel: {\color{\major}Advective} 
    fraction $\tilde{f}_\text{\conv}$ (red), and the ratios of the vertical to the radial components of velocity field $\tilde{f}_\text{vel}$ (green) and magnetic field $\tilde{f}_\text{mag}$ (blue). {\color{\oldminor}The two vertical gray lines indicate the time for Figs. \ref{fig:slice_convective} and \ref{fig:slice_radiative}.}}\label{fig:evolution}
\end{figure}

\begin{figure}
  {\color{\oldmajor}
    \centering
    \includegraphics[scale=0.75]{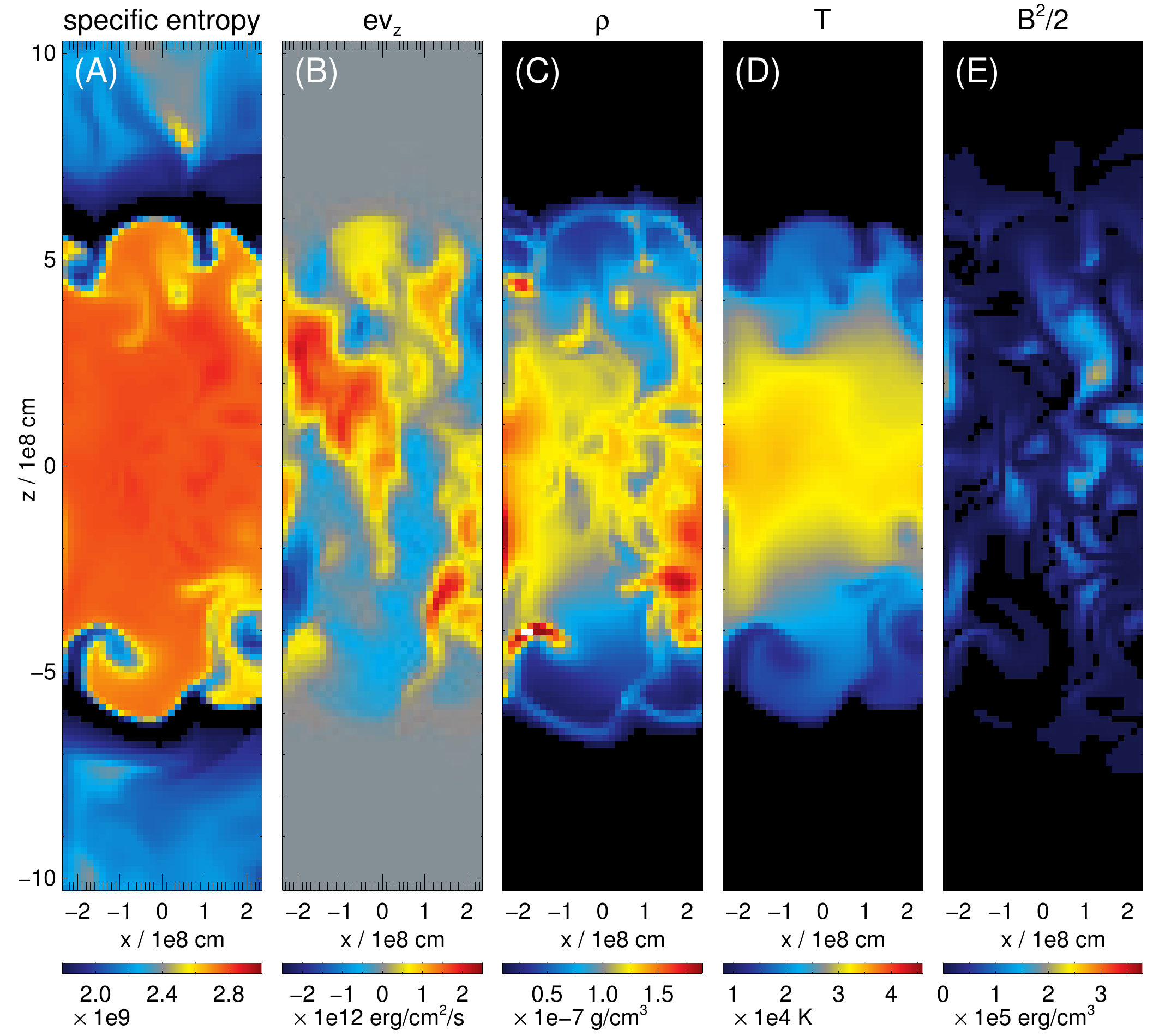}
    \caption{Various quantities on an $x$-$z$ plane ($y = 0$) at $t = 103$ orbits for the case treated in Fig. \ref{fig:evolution} (ws0441F): 
      (A) {\color{\add}specific} entropy, (B) vertical advective heat flux $ev_z$, (C) density $\rho$, (D) gas temperature $T$, and (E) magnetic energy density $\bm{B}^2/2$. Note that images here do not include the entire vertical extent of the box, but instead are limited to the midplane regions.}\label{fig:slice_convective}
  }
\end{figure}

\begin{figure}
{\color{\oldmajor}
    \centering
    \includegraphics[scale=0.75]{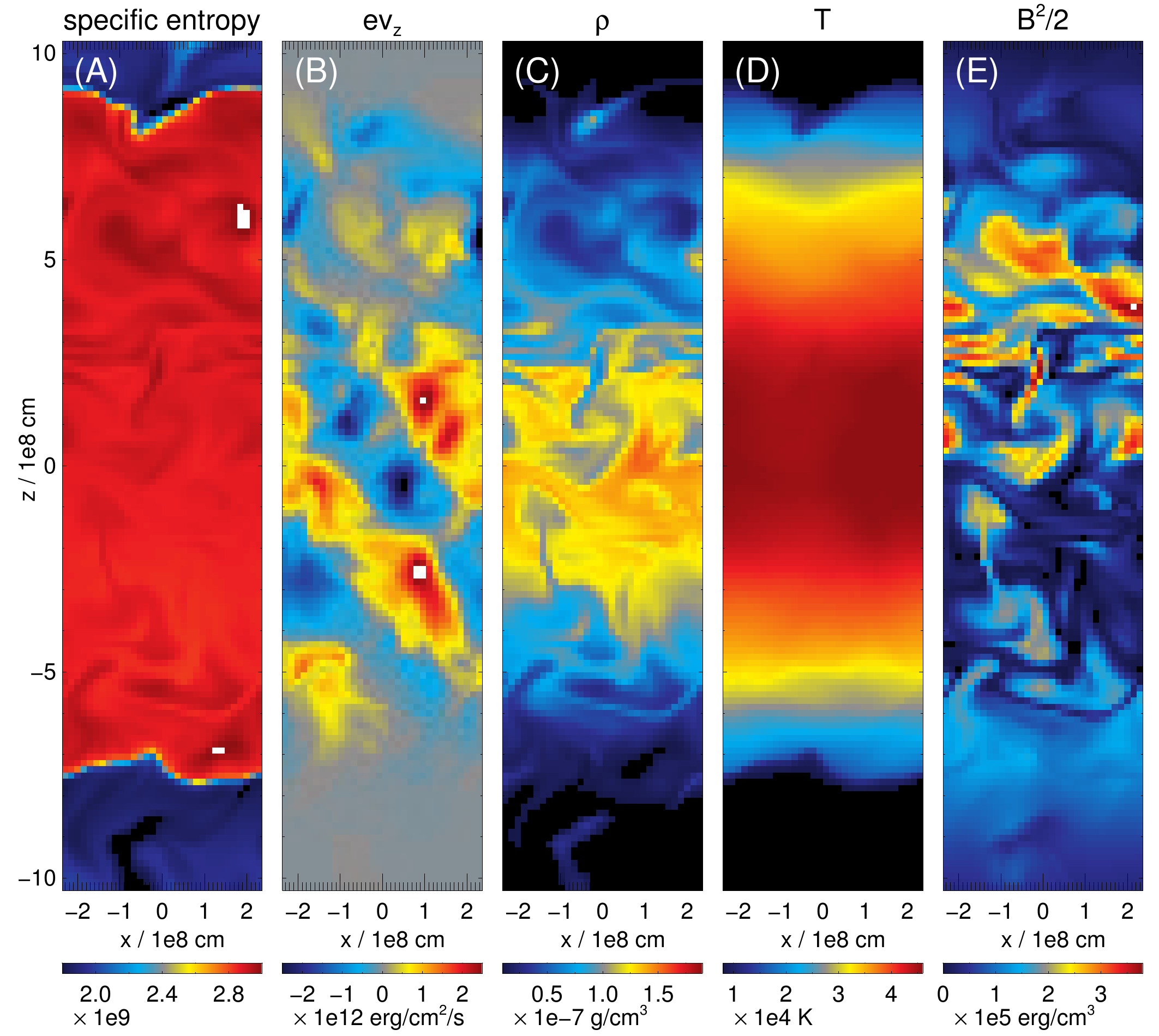}
\caption{The same as Fig. \ref{fig:slice_convective} except at $t = 90$ orbits.}\label{fig:slice_radiative}
}
\end{figure}

\begin{figure}
    \centering
    \includegraphics[scale=0.75]{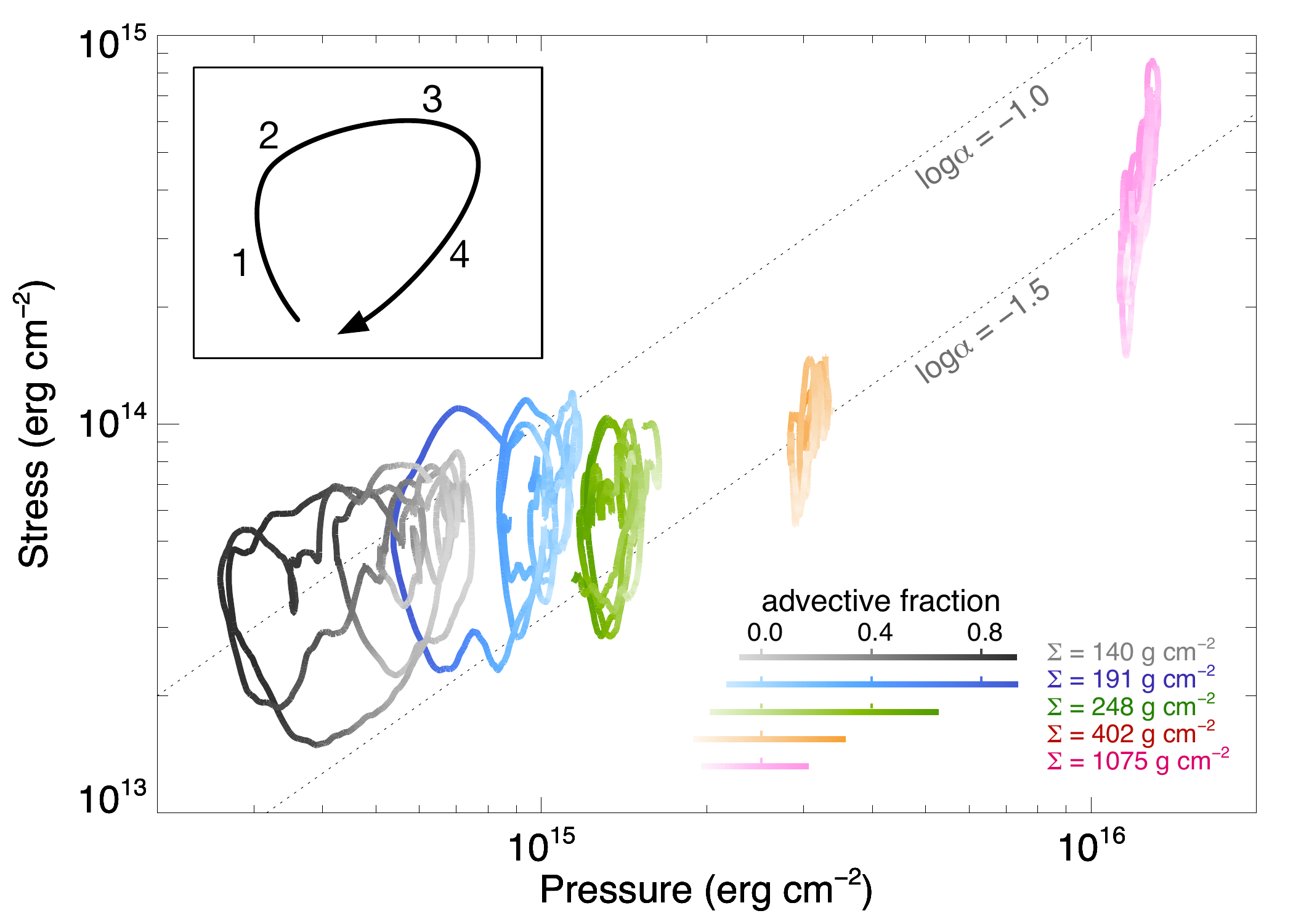}
  \caption{Trajectories in the plane of vertically-integrated pressure $\tilde{P}_\text{thermal}(t)$
  vs. stress $\tilde{W}_{xy}(t)$ for selected simulations on the upper branch ($\Sigma_0$ = 140 {\color{\oldminor}(ws0441F)}, 191 {\color{\oldminor}(ws0427F)},
  248 {\color{\oldminor}(ws0472F)}, 402 {\color{\oldminor}(ws0469F)}, and 1075 g cm$^{-2}$ {\color{\oldminor}(ws0429F)}) indicated by different colors. The color intensity
  represents the {\color{\major}advective} 
  fraction $\tilde{f}_\text{\conv}$. The dotted lines are contours of $\alpha$ =
  0.1 and 0.03. The insert at the upper left shows a schematic picture of the
  limit-cycle described in the text.}\label{fig:limitcycle}
\end{figure}

\begin{figure}
    \centering
    \includegraphics[scale=0.75]{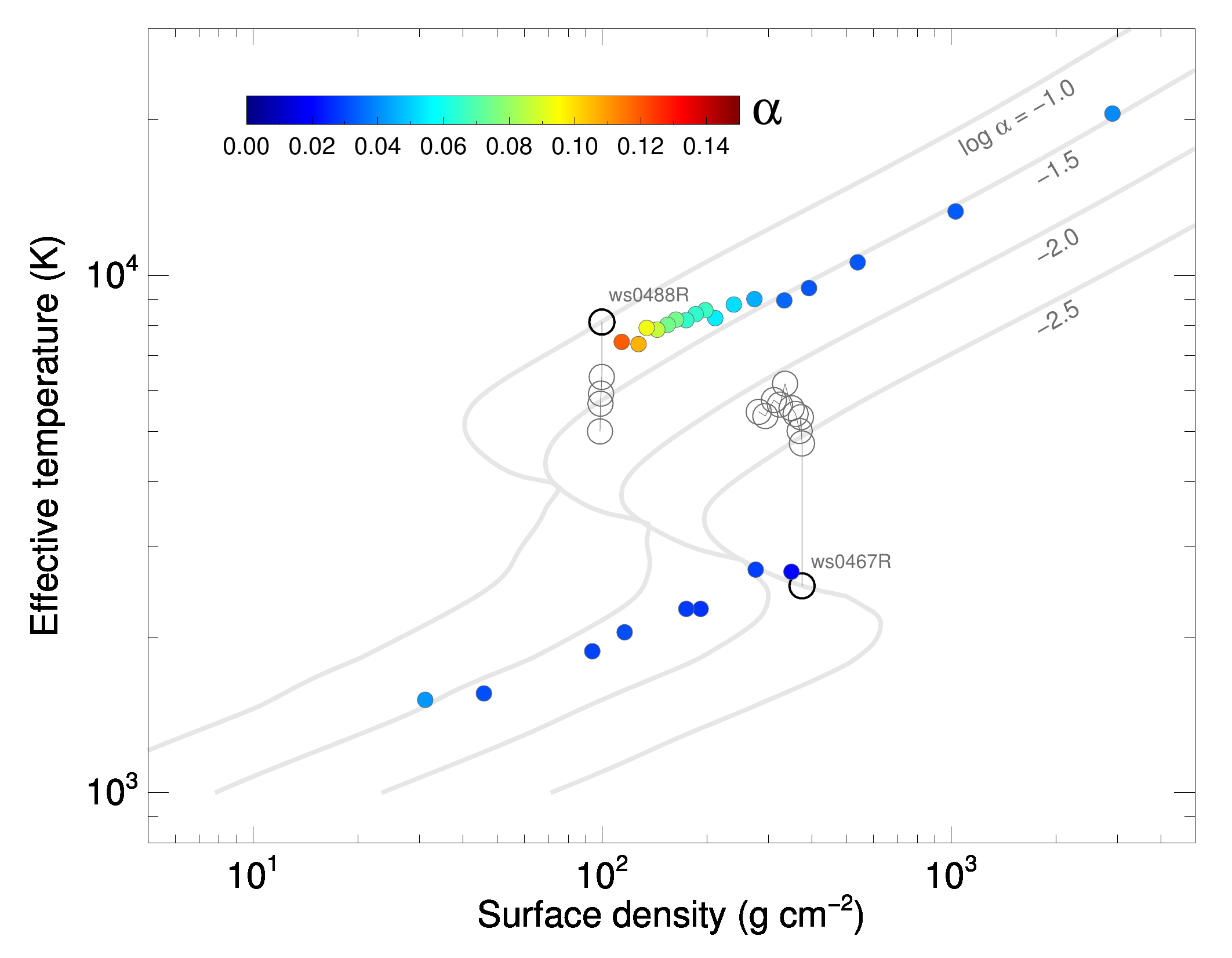}
\caption{Runaway cooling (ws0488R) and heating (ws0467R) beyond the edges of the stable
  branches: Gray thin lines are trajectories from the initial conditions shown as
  black open circles. Gray open circles denote positions every ten orbits on the
  trajectories. See Table \ref{table} for the labels of the runs. Other notations are the same as in Fig. \ref{fig:s-curve}.}\label{fig:edge}
\end{figure}

\begin{figure}
    \centering
    \includegraphics[scale=0.75]{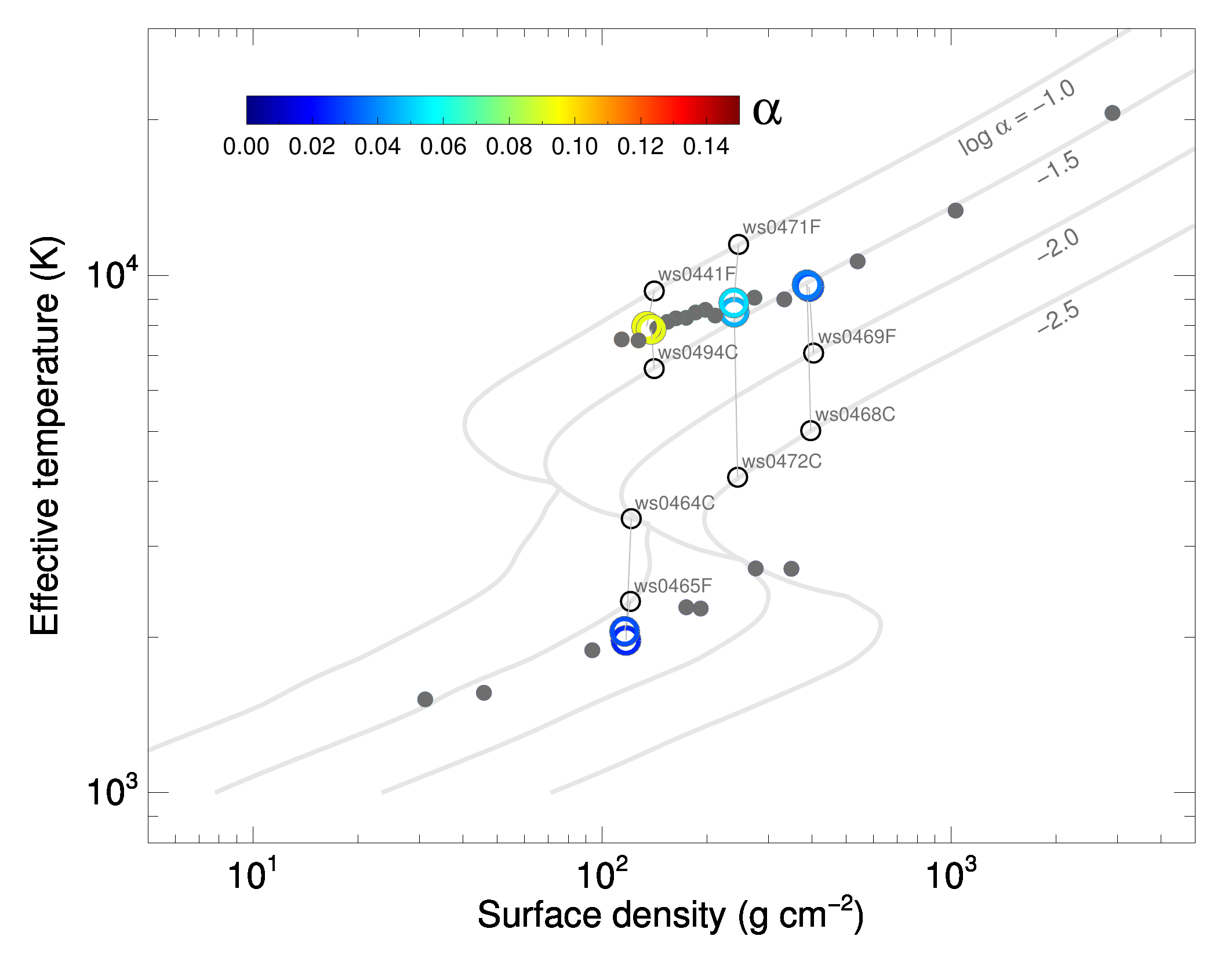}
\caption{Dependence on the initial effective temperature: Runs that
  have almost the same surface density, but have different initial effective
  temperatures are compared. Gray straight lines connect the initial conditions
  (black open circles) and the corresponding final equilibrium states (colored
  open circles). Other notations are the same as in Fig. \ref{fig:edge} except that other
  runs are shown as gray filled circles for clarity.}\label{fig:diff}
\end{figure}

\begin{figure}
    \centering
    \includegraphics[scale=0.75]{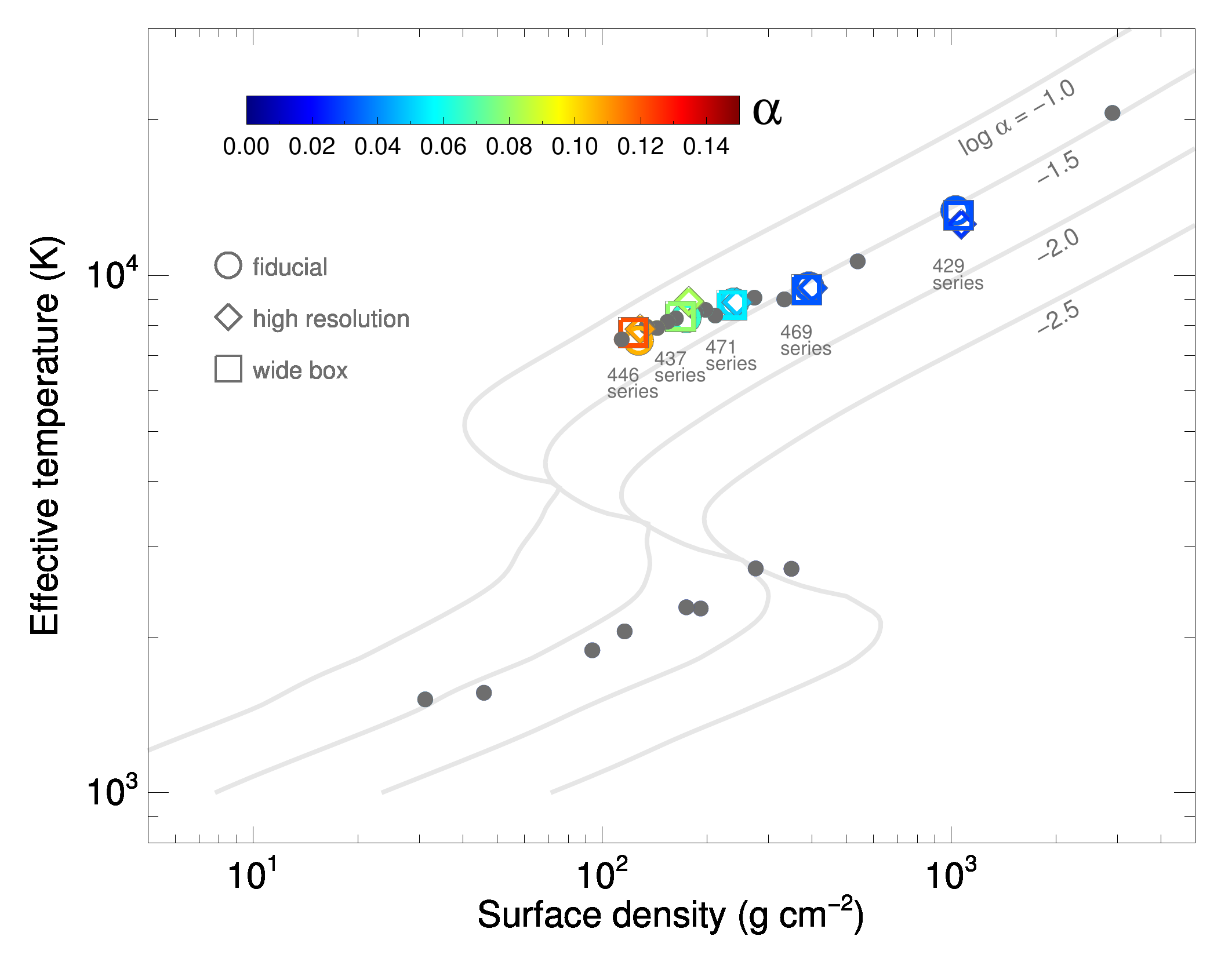}
\caption{Numerical convergence check: for selected surface densities
  ($\Sigma_0 =$ 132, 179, 247, 402, and 1075 g cm$^{-2}$), a high resolution run
  (open diamond), a wide box run (open square), and the fiducial run (open
  circle) are compared in the plane of surface density vs. effective
  temperature. Other notations are the same as in Fig. \ref{fig:edge} except that other
  runs are shown as gray filled circles for clarity.}\label{fig:conv}
\end{figure}

\begin{figure}
{\color{\oldmajor}
    \centering
    \includegraphics[scale=0.45]{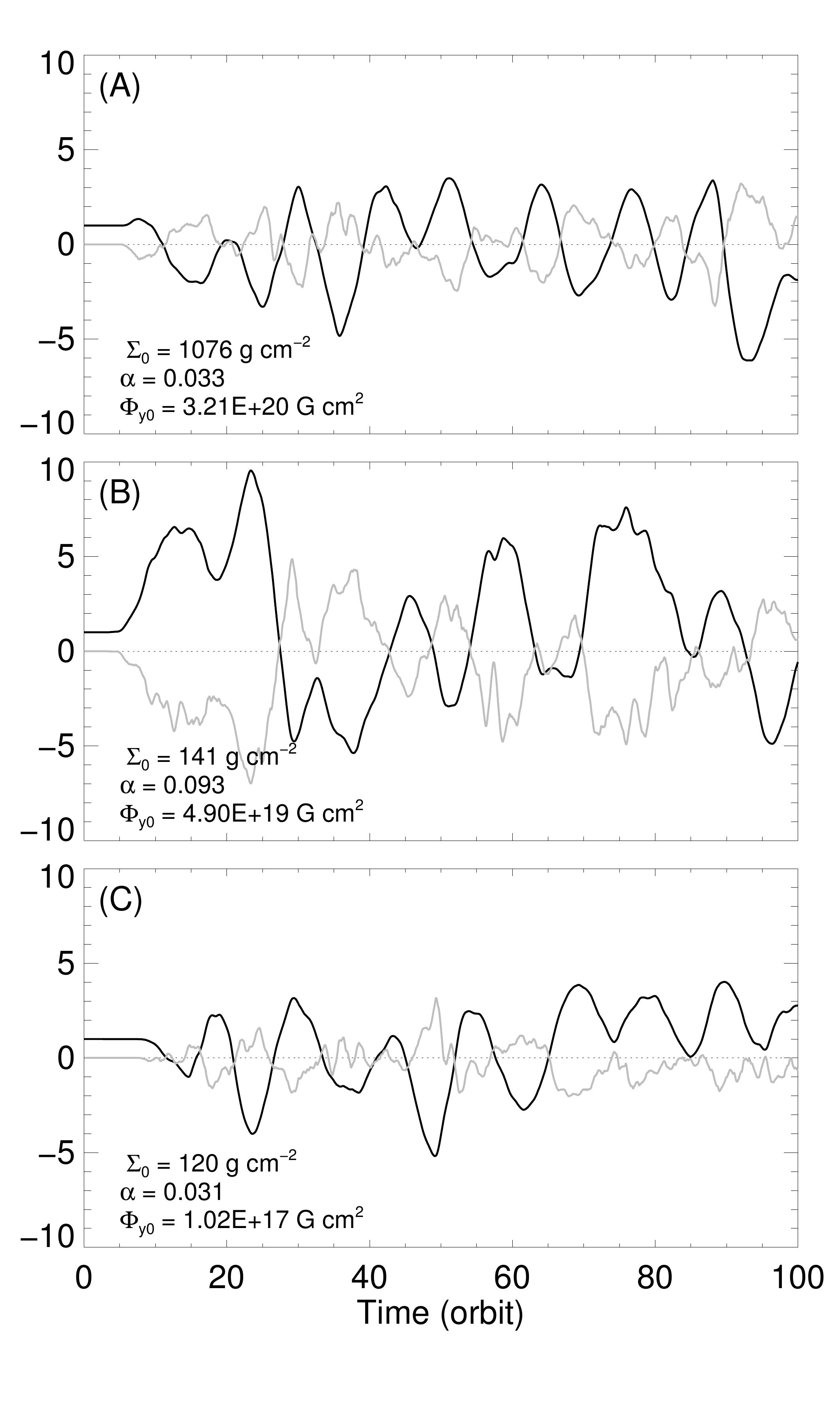}
\caption{Time variation of the net toroidal flux $\tilde{\Phi}_y$ (black) and the net radial flux $\tilde{\Phi}_x\times5$ (gray), each divided by the initial net toroidal flux ${\Phi_y}_0$, for the initial hundred orbits for three selected runs: (A) $\Sigma_0 =$ 1076 (ws0429F), (B) 141 (ws0441F), and (C) 120 (ws0465F) g cm$^{-2}$. The values of $\alpha$ and the initial net toroidal flux ${\Phi_y}_0$ are also shown in each panel.}\label{fig:toroidalflux}
}
\end{figure}







\clearpage

\begin{deluxetable}{lrrrrrrrrrrrrrrrrrrrr}
\tabletypesize{\scriptsize}
\rotate
\tablecaption{List of Runs.}
\tablewidth{0pt}
\tablehead{
\colhead{run} & \colhead{$\Sigma_0$} & \colhead{${T_\text{eff}}_0$} & \colhead{$\alpha_0$} & \colhead{$h_0$} &
\colhead{$\beta_0$} & \colhead{$\bar{\Sigma}$} & \colhead{$\bar{T}_\text{eff}$} &
\colhead{$\alpha$} & \colhead{$\tau_\text{tot}$} & \colhead{$N_x$} & \colhead{$N_y$} & \colhead{$N_z$} &\colhead{$L_x/h_0$} & \colhead{$L_y/h_0$} & \colhead{$L_z/h_0$} &
\colhead{$L_z/h_\text{p}$} & \colhead{$t_\text{th}$} & \colhead{$t_1$} & \colhead{$t_2$}
}
\startdata
\cutinhead{Upper branch solutions}
ws0430F & 2983 & 14454 & 0.0097 & 1.81E+09 & 100 & 2900 & 20617 & 0.0401 & 37931 & 32 & 64 & 256 & 0.500 & 2.000 & 4.000 & 8.37 & 6.53  & 50 & 150 \\
ws0429F & 1075 & 10000 & 0.0098 & 1.41E+09 & 100 & 1030 & 13352 & 0.0332 & 24232 & 32 & 64 & 256 & 0.500 & 2.000 & 4.000 & 8.54 & 9.49 & 90 & 190 \\
wa0429H & 1075 & 10000 & 0.0098 & 1.41E+09 & 100 & 1070 & 12570 & 0.0279 & 29083 & 48 & 96 & 384 & 0.600 & 2.400 & 4.800 & 10.4 & 11.5 & 90 & 190 \\
wv0429W & 1075 & 10000 & 0.0098 & 1.41E+09 & 100 & 1052 & 13067 & 0.0311 & 26674 & 64 & 128 & 256 & 1.000 & 4.000 & 4.000 & 8.58 & 10.2 & 20 & 120 \\
ws0439F & 555 & 7943 & 0.0099 & 1.12E+09 & 100 & 540 & 10649 & 0.0325 & 20337 & 32 & 64 & 256 & 0.500 & 2.000 & 4.000 & 7.79 & 10.9 & 50 & 150 \\
ws0469F & 402 & 7079 & 0.0098 & 9.74E+08 & 100 & 391 & 9514 & 0.0323 & 19742 & 32 & 64 & 256 & 0.560 & 2.240 & 4.480 & 8.14 & 11.6  & 50 & 150 \\
wa0469H & 402 & 7079 & 0.0098 & 9.74E+08 & 100 & 400 & 9463 & 0.0318 & 21031 & 48 & 96 & 384 & 0.672 & 2.688 & 5.376 & 9.75 & 11.7 & 50 & 150 \\
wv0469W & 402 & 7079 & 0.0098 & 9.74E+08 & 100 & 385 & 9386 & 0.0328 & 22161 & 64 & 128 & 256 & 1.120 & 4.480 & 4.480 & 8.23 & 11.5 & 50 & 150 \\
ws0468C & 397 & 5011 & 0.0031 & 7.40E+08 & 100 & 386 & 9594 & 0.0375 & 22518 & 32 & 64 & 256 & 0.750 & 3.000 & 6.000 &  8.37 & 10.2 & 50 & 150 \\
ws0491F & 342 & 12882 & 0.0988 & 1.27E+09 & 100 & 332 & 8992 & 0.0347 & 22712 & 32 & 64 & 256 & 0.500 & 2.000 & 4.000 & 9.77 & 11.3 & 50 & 150 \\
ws0470F & 281 & 12022 & 0.0991 & 1.21E+09 & 100 & 273 & 9059 & 0.0452 & 17420 & 32 & 64 & 256 & 0.500 & 2.000 & 4.000 & 9.81 & 8.74 & 50 & 150 \\
ws0472C & 248 & 4073 & 0.0031 & 5.15E+08 & 100 & 239 & 8492 & 0.0459 & 23321 & 32 & 64 & 256 & 1.140 & 4.560 & 9.120 & 10.4 & 9.06 & 50 & 150 \\
ws0471F & 247 & 11481 & 0.0988 & 1.18E+09 & 100 & 238 & 8853 & 0.0522 & 18117 & 32 & 64 & 256 & 0.500 & 2.000 & 4.000 & 10.0 & 7.84 & 50 & 150 \\
wa0471H & 247 & 11481 & 0.0988 & 1.18E+09 & 100 & 243 & 8872 & 0.0496 & 29032 & 48 & 96 & 384 & 0.600 & 2.400 & 4.800 & 12.0 & 8.06 & 20 & 120 \\
wv0471W & 247 & 11481 & 0.0988 & 1.18E+09 & 100 & 235 & 8766 & 0.0546 & 23992 & 64 & 128 & 256 & 1.000 & 4.000 & 4.000 & 10.3 & 7.69 & 50 & 150 \\
ws0492F & 217 & 10964 & 0.0986 & 1.14E+09 & 100 & 211 & 8366 & 0.0548 & 23980 & 32 & 64 & 256 & 0.500 & 2.000 & 4.000 & 10.5 & 7.90 & 50 & 150 \\
ws0425F & 204 & 10715 & 0.0985 & 1.11E+09 & 100 & 197 & 8582 & 0.0668 & 20029 & 32 & 64 & 256 & 0.500 & 2.000 & 4.000 & 10.1 & 6.48 & 50 & 150 \\
ws0427F & 191 & 10471 & 0.0981 & 1.09E+09 & 100 & 185 & 8482 & 0.0646 & 16517 & 32 & 64 & 256 & 0.500 & 2.000 & 4.000 & 10.0 & 6.66 & 50 & 150 \\
ws0437F & 179 & 10232 & 0.0975 & 1.07E+09 & 100 & 174 & 8283 & 0.0651 & 16791 & 32 & 64 & 256 & 0.500 & 2.000 & 4.000 & 10.1 & 6.64 & 50 & 150 \\
wa0437H & 179 & 10232 & 0.0975 & 1.07E+09 & 100 & 177 & 8908 & 0.0821 & 14401 & 48 & 96 & 384 & 0.600 & 2.400 & 4.800 & 11.5 & 5.17 & 50 & 150 \\
wv0437W & 179 & 10232 & 0.0975 & 1.07E+09 & 100 & 168 & 8333 & 0.0795 & 23234 & 64 & 128 & 256 & 1.000 & 4.000 & 4.000 & 10.6 & 5.79 & 50 & 150 \\
ws0433F & 168 & 10000 & 0.0978 & 1.05E+09 & 100 & 162 & 8263 & 0.0749 & 15595 & 32 & 64 & 256 & 0.500 & 2.000 & 4.000 & 10.2 & 5.99 & 50 & 150 \\
ws0436F & 158 & 9772 & 0.0978 & 1.03E+09 & 100 & 154 & 8138 & 0.0747 & 17608 & 32 & 64 & 256 & 0.500 & 2.000 & 4.000 & 10.3 & 6.05 & 30 & 130 \\
wt0487F & 149 & 9549 & 0.0983 & 1.00E+09 & 100 & 143 & 7914 & 0.0862 & 20503 & 32 & 64 &256 & 0.500 & 2.000 & 4.000 & 10.6 & 5.58 & 50 & 150 \\
ws0441F & 140 & 9332 & 0.0981 & 9.77E+08 & 100 & 134 & 7966 & 0.0927 & 17556 & 32 & 64 & 256 & 0.500 & 2.000 & 4.000 & 10.2 & 5.12 & 50 & 150 \\
ws0494C	& 140 &	6606 & 0.0311 & 7.07E+08 & 10  & 138 & 7868 & 0.0919 & 23865 & 32 & 64 & 256 & 0.700 & 2.800 & 5.600 & 10.8 & 5.29 & 10 & 90 \\
ws0446F & 132 & 9120 & 0.0987 & 9.51E+08 & 100 & 127 & 7490 & 0.1062 & 34556 & 32 & 64 & 256 & 0.500 & 2.000 & 4.000 & 12.0 & 5.06 &50 & 150 \\
wa0446H & 132 & 9120 & 0.0987 & 9.51E+08 & 100 & 128 & 7876 & 0.1087 & 23795 & 48 & 96 & 384 & 0.600 & 2.400 & 4.800 & 13.0 & 4.66 &50 & 150 \\
wz0446W & 132 & 9120 & 0.0987 & 9.51E+08 & 10 & 123 & 7758 & 0.1210 & 28978 & 64 & 128 & 256 & 1.000 & 4.000 & 4.000 & 11.4 & 4.35 &25 & 125 \\
wt0442F & 118 & 8709 & 0.0982 & 8.97E+08 & 10 & 113 & 7522 & 0.1195 & 22037 & 32 & 64 & 256 & 0.500 & 2.000 & 4.000 & 11.4 & 4.53 &50 & 150 \\
ws0488R & 99 & 8128 & 0.0982 & 8.12E+08 & 10 & --- & --- & --- & --- & 32 &64 &256 & 0.500 & 2.000 & 4.000 & --- & --- &--- & --- \\
\cutinhead{Lower branch solutions}
ws0467R & 373 & 2511 & 0.0031 & 1.24E+08 & 100 & --- & --- & --- & --- & 32 & 64 & 256 & 1.000 & 4.000 & 8.000 & ---  & ---& --- & --- \\
{\color{\oldmajor}ws0466F} & 353 & 1584 & 0.0031 & 8.18E+07 & 100 & 348 & 2675 & 0.0204 &13.2 & 32 & 64 & 256 & 0.750 & 3.000 & 6.000 & 8.18 & 17.3 & 20 & 100 \\
ws0438F & 285 & 1096 & 0.0010 & 1.13E+08 & 100 & 275 & 2714 & 0.0287 & 8.8 & 32 & 64 & 256 & 0.500 & 2.000 & 4.000 & 7.70 & 12.3 & 50 & 150 \\
wt0435F & 211 & 1000 & 0.0010 & 1.06E+08 & 100 & 191 & 2271 & 0.0271 & 2.6 & 32 & 64 & 256 & 0.450 & 1.800 & 3.600 & 7.44 & 10.9 & 120 & 220 \\
ws0462F & 176 & 1778 & 0.0099 & 6.51E+07 & 100 & 174 & 2283 & 0.0282 & 2.4 & 32 & 64 & 256 & 1.000 & 4.000 & 8.000 & 10.1 & 9.78 & 50 & 150 \\
ws0464C & 121 & 3388 & 0.0313 & 1.35E+08 & 100 & 117 & 1969 & 0.0254 & 2.7 & 32 & 64 & 256 & 0.500 & 2.000 & 4.000 & 10.9 & 10.6 & 50 & 150 \\
ws0465F & 120 & 2344 & 0.0314 & 5.18E+07 & 100 & 116 & 2051 & 0.0312 & 2.3 & 32 & 64 & 256 & 1.000 & 4.000 & 8.000 & 8.29 & 8.89 & 50 & 150 \\
ws0434F & 100 & 1513 & 0.0099 & 5.85E+07 & 100 & 93 & 1885 & 0.0295 & 3.2 & 32 & 64 & 256 & 0.750 & 3.000 & 6.000 & 7.12 & 9.75 & 50 & 150 \\
ws0476F & 49 & 1698 & 0.0315 & 5.28E+07 & 100 & 45 & 1560 & 0.0314 & 5.4 & 32 & 64 & 256 & 0.800 & 3.200 & 6.400 & 7.34 & 9.02 & 50 & 150 \\
ws0445F & 32 & 1096 & 0.0098 & 8.63E+07 & 100 & 31 & 1515 & 0.0418 & 4.4 & 32 & 64 & 256 & 0.500 & 2.000 & 4.000 & 7.52 & 6.66 & 50 & 150 \\
\enddata
\tablecomments{The last letters in the names of runs denote the following: F = fiducial run, H =
  high resolution run, W = wide box run, R = runaway heating/cooling run, and C
  = run to check dependence on the initial condition. The units of surface
  densities ($\Sigma_0$ and $\bar{\Sigma}$), effective temperatures
  (${T_\text{eff}}_0$ and $\bar{T}_\text{eff}$), height ($h_0$), and thermal time ($t_\text{th}$) are, respectively, g cm$^{-2}$, K, cm, and orbit. $L_x$, $L_y$,
  and $L_z$ are the lengths, and $N_x$, $N_y$, and $N_z$ are the numbers of cells,
  in the $x$, $y$, and in $z$ directions, respectively. 
  The time-averaging in diagnostics is done for $t_1 < t < t_2$ orbits. 
  The pressure scale height of the steady state is computed as $h_\text{p} \equiv \int\left[\left<p_\text{thermal}\right>\right]dz / 2\max(\left[\left<p_\text{thermal}\right>\right]) $.}\label{table}
\end{deluxetable}





\end{document}